# The role of spin-orbit coupling and state-crossing topography in the non-radiative decay of Ir(III) complexes

Iván Soriano-Díaz,[1] Ilya D. Dergachev,[2] Sergey A. Varganov,[3] Enrique Ortí,[1] and Angelo Giussani[1,*]

*[1] Institute for Molecular Science (ICMol), Universitat de València, Catedrático José Beltrán 2, 46980 Paterna, España.*

*[2] Department of Chemistry, University of Nevada, Reno, 1664 N. Virginia Street, Reno, NV 89557-0216, USA; Current address: Department of Chemistry, New York University, New York, New York 10003, USA*

*[3] Department of Chemistry, University of Nevada, Reno, 1664 N. Virginia Street, Reno, NV 89557-0216, USA*

*A pillar of our current understanding of the photoluminescence of Ir(III) complexes is the assumption that the population of triplet metal-centered states determines an efficient non-radiative decay to the ground state minimum. Based on that assumption, the energy separation between the emitting state and the minimum-energy crossing point of the triplet metal-centered and the ground states has been employed as a key variable for evaluating the ability of Ir(III) complexes to decay non-radiatively. We demonstrate that the strong spin-orbit coupling between the triplet metal-centered and the ground state of Ir(III) complexes, together with the sloped topography of their crossing, lead to a significant energy separation between the two states, resulting in a reduced rate of non-radiative ground state recovery. Therefore, we propose that the role of metal-centered states is defined by the tendency of the excited state population to remain trapped in the metal-centered minima.*

Ir(III) transition-metal complexes (Ir3-TMCs),[1–8] which are characterized by a $d^6$ electronic configuration and strong spin-orbit coupling (SOC), are fundamental for electroluminescent devices and photodynamic therapy (see Section S1-S2 of the Supporting Information). Because for both applications a slow non-radiative decay is required, much effort has been devoted to determining the main non-radiative processes at play, with the ultimate goal of devising chemical modifications that minimize them. In general, two intramolecular non-radiative decays are considered.[9–11] The first is the intersystem crossing (ISC) from the emitting triplet minimum to the ground state ($S_0$). The second is the population of triplet metal-centered states ($^3$MC), which are expected to efficiently mediate a non-radiative decay to the ground state.

$^3$MC states of Ir3-TMCs are characterized by equilibrium structures in which at least one of the coordination metal bonds is dissociated (Section S2). This dissociation causes a massive rise of the ground state energy at the $^3$MC minima and the presence of energetically and geometrically nearby $^3$MC/$S_0$ minimum-energy crossing points (MECPs).[12–16] It is the existence of such accessible MECPs, that, together with the large SOC, leads to the assumption that $^3$MC states efficiently drive the population back to $S_0$. Most of the theoretical understanding of the photophysics of Ir3-TMCs is based on the validity of this assumption, relating an increase/decrease of the emission quantum yield with a decrease/increase in accessibility of $^3$MC states.

The presumed highly efficient decay of $^3$MC states, has led the scientific community to employ transition state theory (TST) to characterize the $^3$MC mediated non-radiative decay, treating the $^3$MC/$S_0$ MECP as an analog of the transition state (TS) geometry.[9–11,14,17] This implies that the probability of decay to $S_0$ is equal to unity at MECP, as in TST the probability of reaction is unity once the system reaches a TS. Here we demonstrate that for Ir3-TMCs the idea that $^3$MC/$S_0$ MECP mediates an efficient re-population of the Frank-Condon region can not be assumed to be in general valid.

As a representative example, we focus on $[Ir(ppy)_2(bpy)]^+$ (where Hppy is 2-phenylpyridine and bpy is 2,2′-bipyridine), which is an archetype of the $[Ir(C^\wedge N)_2(N^\wedge N)]^+$ cyclometalated Ir(III) complexes used in electroluminescent applications.[2,3] Our results are however general and apply to any Ir3-TMCs. The emitting $T_1$ state of $[Ir(ppy)_2(bpy)]^+$ has metal-ligand charge transfer (MLCT) character, and displays an equilibrium structure similar to the $S_0$ minimum.[13,18] From $T_1$, the processes determining the efficiency of light emission are supposed to start: phosphorescent emission, direct ISC to $S_0$, and population of the $^3$MC states with the subsequent transfer back to $S_0$ via the $^3MC/S_0$ MECP. The latter process is a $T_1$-to-$S_0$ decay mediated by the $^3MC/S_0$ MECP, because, even if the nature of the triplet state changes from MLCT to MC, the involved potential energy surface (PES) is always the adiabatic $T_1$. Two types of $^3$MC states have been identified in $[Ir(ppy)_2(bpy)]^+$.[13] The axial $^3MC_{ax}$ state is characterized by a minimum structure $(^3MC_{ax})_{min}$ displaying the dissociation of one Ir-$N_{ppy}$ bond, and in the equatorial $^3MC_{eq}$ state, a Ir-$N_{bpy}$ bond is broken in its equilibrium geometry $(^3MC_{eq})_{min}$. Both bonds are present in the octahedral structures of the emitting $T_1$ MLCT and $S_0$ minima, so the evolution from the former to the latter, through the $^3MC/S_0$ MECP, implies the stretching of an Ir-N bond up to dissociation and then the re-formation of the very same bond (Figure 1). Therefore, an Ir-N distance plays the role of the reaction coordinate connecting the $T_1$ minimum and the $^3MC/S_0$ MECP. This means that both the initial and final minima (the emitting $T_1$ and the $S_0$ minima) are located along the same direction relative to the MECP. Borrowing from the terminology associated with conical intersections (CIs), we can say that the $^3MC/S_0$ intersection is sloped, with the gradients of two states pointing toward the same direction. [19–23]

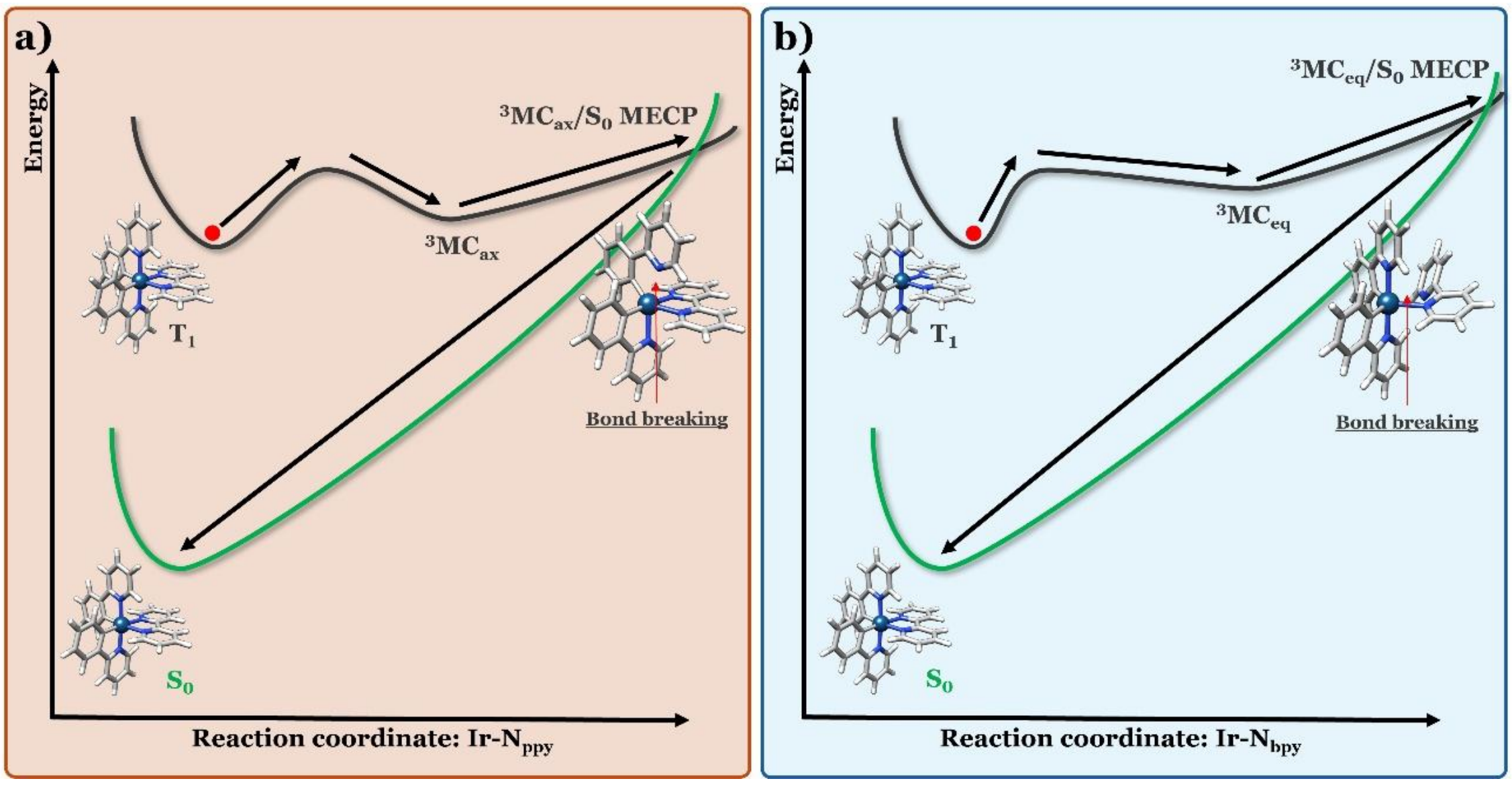

**Figure 1.** Non-radiative decay of $[Ir(ppy)_2(bpy)]^+$ from the $T_1$ minimum to the $S_0$ minimum mediated by (a) the $^3MC_{ax}/S_0$ MECP, and (b) the $^3MC_{eq}/S_0$ MECP.

The topography of the MECP and the value of the SOC between the $^3$MC and $S_0$ states are key variables that strongly determine the probability of non-radiative decay. To illustrate this most clearly, we first need to look at the theory behind non-adiabatic events (NAEs, Section S3).[21,24–26] Let us first analyze the case of an NAE between two adiabatic electronic states of pure spin having the same spin multiplicity (i.e. an internal conversion, IC). Expanding the wave-function in the basis of these two spin-pure adiabatic electronic states, it is possible to show that what causes the passage of population from one PES to the other are the non-adiabatic couplings, NACs (eq. S8 and 1). Since, according to the generalized Hellman-Feynman theorem, the value of the derivative coupling between two states, which is the main component of the corresponding NAC, is inversely proportional to their energy separation (eq. S9 and 2), the closer are the states, the more probable is the NAE, and that is why CIs are key in photochemistry.[19,22]

In the case of considering just the $S_0$ and first singlet excited state, $S_1$, the discussed equations will read as follows.

$$i\dot{\psi}^{S_1}(\boldsymbol{R},t) = \left(\hat{T}_R + V^{S_1}(\boldsymbol{R})\right)\psi^{S_1}(\boldsymbol{R},t) + \hat{\Lambda}^{S_0S_1}\psi^{S_0}(\boldsymbol{R},t) \quad eq.1$$

$$\hat{\Lambda}^{S_0S_1} \cong \langle S_0|\hat{\nabla}_R|S_1\rangle = \frac{\langle S_0|\hat{\nabla}_R\hat{H}_{el}|S_1\rangle}{V^{S_1}(\boldsymbol{R}) - V^{S_0}(\boldsymbol{R})} \quad eq.2$$

Where $\dot{\psi}^{S_1}(\boldsymbol{R},t)$ is the time-derivative of the nuclear wave-function associated with state $S_1$, $\psi^{S_0}(\boldsymbol{R},t)$ is the nuclear wave-function associated with $S_0$, $\hat{\Lambda}^{S_0S_1}$ is the NAC between $S_0$ and $S_1$, $V^{S_1}(\boldsymbol{R})$ and $V^{S_0}(\boldsymbol{R})$ are respectively the $S_1$ and $S_0$ PESs, $\langle S_0|\hat{\nabla}_R|S_1\rangle$ is the derivative coupling between $S_0$ and $S_1$, and $\hat{T}_R$ is the nuclear kinetic energy operator.

Let us now expand our wave function in the basis of two spin-pure adiabatic electronic states having different spin multiplicities, for example, a triplet and a singlet state. Such states are also called spin-diabatic states.[27–31] Repeating the previous mathematical manipulations, it now appears that the corresponding NAC is zero due to the orthogonality of the different spin wave functions, and that what couples the two states and drives a NAE is the SOC term (see eq. 3), assuming it is included in the Hamiltonian.[32–34] Considering only the $S_0$ and $T_1$ states, the coupling equation can be written as follows.

$$i\dot{\psi}^{T_1}(\boldsymbol{R},t) = \left(\hat{T}_R + V^{T_1}(\boldsymbol{R})\right)\psi^{T_1}(\boldsymbol{R},t) + H_{SOC}^{S_0T_1}\psi^{S_0}(\boldsymbol{R},t) \quad eq.3$$

Where $\dot{\psi}^{T_1}(\boldsymbol{R},t)$ is the time-derivative of the nuclear wave-function associated with state $T_1$, and $\hat{H}_{SOC}^{S_0T_1}$ is the SOC between $S_0$ and $T_1$.

Let us again use the two spin-pure adiabatic electronic states of different spin symmetry (a singlet and a triplet) as a basis for the expansion of the wave function, but instead of using them as they are, let us use the combination resulting from the diagonalization of the electronic Hamiltonian including the SOC term (see eq. 4).

$$\begin{pmatrix} V^{S_0} & H_{SOC}^{S_0T_1} \\ H_{SOC}^{S_0T_1} & V^{T_1} \end{pmatrix} \xrightarrow{diagonalization} \begin{pmatrix} V^{SM_0} & 0 \\ 0 & V^{SM_1} \end{pmatrix} \quad eq.4$$

We call the resulting states, no longer of pure spin multiplicity, spin-mixed or spin-adiabatic states. The energy separation between these spin-adiabatic states depends on the magnitude of the SOC between the original spin-diabatic states. If we expand the wave function in the basis of such spin-adiabatic states, they will be again coupled by NAC, which determines a NAE, now computed over the two spin-mixed states (see eq. 5). Again, the value of the NAC is inversely proportional to the interstate energy separation, therefore the closer are the spin-mixed states, the more probable is the NAE. It must be remembered that the NAC is now different from zero only because of the mixing of the original singlet and triplet states caused by their SOC. The equation coupling the two spin-mixed states $SM_0$ and $SM_1$, resulting from the mixing of the original spin-pure states $S_0$ and $T_1$, will be as follows.

$$i\dot{\psi}^{SM_1}(\boldsymbol{R},t) = \left(\hat{T}_R + V^{SM_1}(\boldsymbol{R})\right)\psi^{SM_1}(\boldsymbol{R},t) + \hat{\Lambda}^{SM_0SM_1}\psi^{SM_0}(\boldsymbol{R},t) \quad eq.5$$

Spin-pure states and spin-mixed states are two complete basis sets spanning the same space, so the description of Ir3-Ru2-TMCs using either representation must be equivalent. Let us first describe $[Ir(ppy)_2(bpy)]^+$ using spin-mixed states. At MECP, the energy separation between the latter states can be estimated as twice the SOC between the corresponding spin-pure states. The computed SOC (Section S4) between the $^3MC_{ax}$ and $S_0$ states at the $(^3MC_{ax})_{min}$ structure is 3400 $cm^{-1}$ (0.42 eV), and for the $^3MC_{eq}$ and $S_0$ states at $(^3MC_{eq})_{min}$, is 1676 $cm^{-1}$ (0.21 eV). At the corresponding $^3MC_{ax}/S_0$ and $^3MC_{eq}/S_0$ MECPs, the splitting of the original spin-pure $^3MC$ and $S_0$ states when including the SOC is 0.84 and 0.42 eV.

It is fundamental to understand that, depending on the relative position of the initial and final minima with respect to MECP, the SOC-induced splitting of the $^3MC$ and $S_0$ states has a completely different effect on the decay process. When the initial and final minima are located in opposite directions relative to MECP (i.e., when it is possible to define a reaction coordinate along which the reactant minimum, the MECP, and the product minimum are consecutive points) the MECP has a peaked topography.[19–21] This case has been the subject of different studies.[28,35–37] The mixing induced by the SOC results in two separated spin-adiabatic states, and what in the spin-pure picture was described as an ISC (Figure 2a, upper panel) now instead corresponds to an adiabatic process, in which the system evolves from one minimum to another along the same spin-adiabatic PES via TS barrier (Figure 2a, bottom panel). When instead the initial and final minima are located in the same directions relative to MECP, the MECP has a sloped topography.[19–21] This case,[38–41] is the one for the $T_1$-to-$S_0$ decay mediated by $^3MC$ states in Ir3-Ru2-TMCs. The mixing induced by the SOC again results in two separated spin-adiabatic states, but what in the spin-pure picture was described as an ISC now still corresponds to a NAE (Figure 2b). Such NAE is driven by the NAC between the spin-mixed states that at the original MECP are separated by a value equal to two times their SOC. Consequently, it is a NAE whose probability has no reason to be equal to unity at MECP, in turn meaning that the probability of reaching the $S_0$ minimum has no reason to be high. In our example, according to the computed SOCs, at the $^3MC_{ax}/S_0$ and $^3MC_{eq}/S_0$ MECPs, the spin-mixed states are separated by as much as 0.84 and 0.42 eV, respectively.

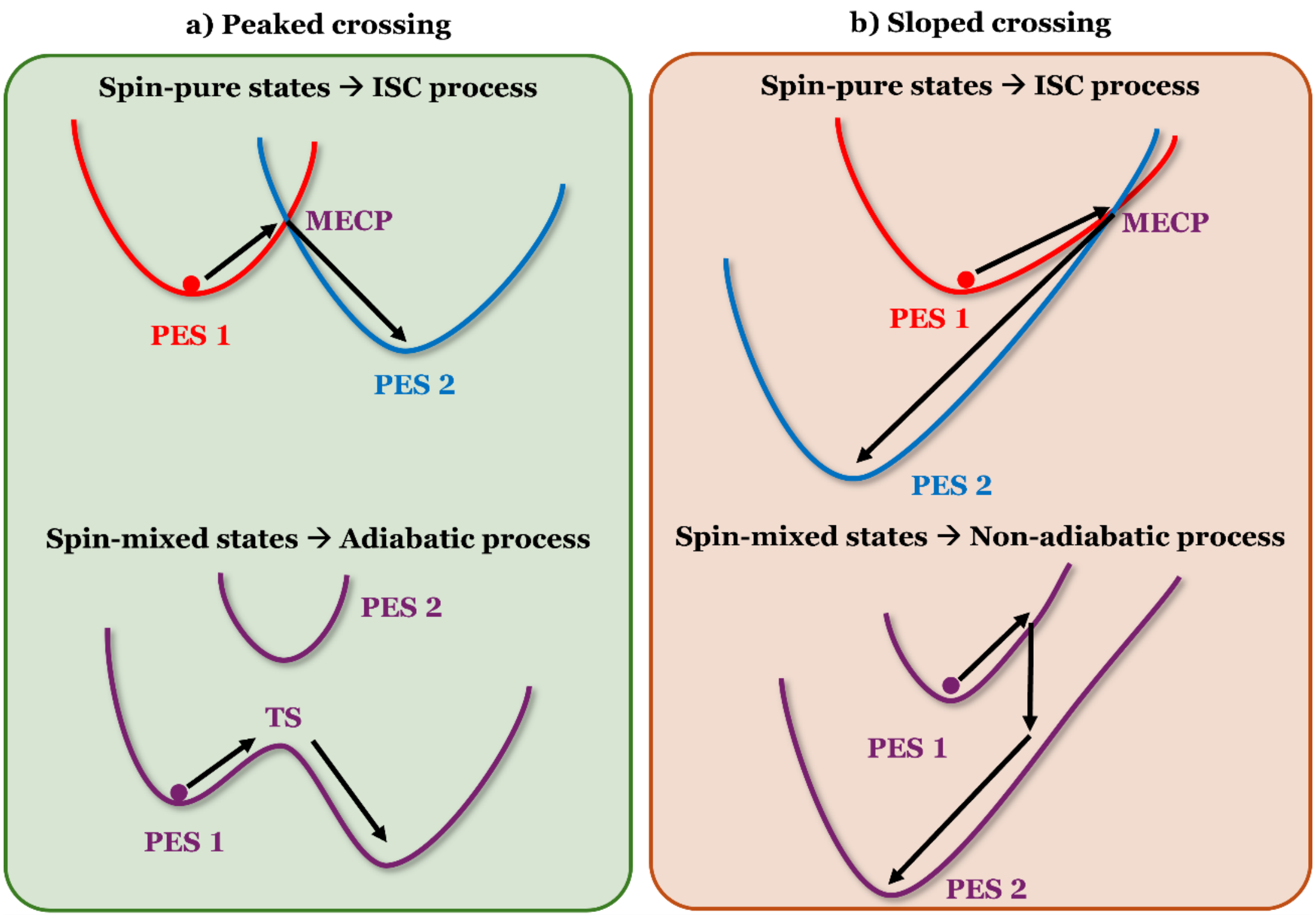

**Figure 2.** Spin-pure and spin-mixed representations of peaked (a) and sloped (b) MECP-mediated processes.

The same conclusion is also obtained using the spin-pure representation (Figure 2b, upper panel). Once the trajectory propagating on PES1 reaches MECP, the large SOC results in a high (close to unity) probability of NAE. But due to the direction of propagation, it is the high-energy branch of PES2 that becomes populated. Once the system runs out of kinetic energy, it goes back to MECP, where again the large SOC implies a high NAE, now to the low-energy branch of PES1. This results in a very slow population decay to the low-energy branch of PES2, meaning a low tendency to re-populate the $S_0$ Franck-Condon region. This conclusion is confirmed by computing the corresponding probability at the $^3MC_{eq}/S_0$ MECP of $[Ir(ppy)_2(bpy)]^+$ using non-adiabatic TST, NA-TST, implemented in the NAST code (Section S5).[28,42]

Using the NA-TST, it is possible to evaluate the rate constant associated with the $^3$MC-mediated non-radiative decay of $[Ir(ppy)_2(bpy)]^+$. An agreement with the experimental data would constitute an important piece of evidence for the here-presented model in which the probability of non-radiative re-population of the ground state is much lower than unity. However, such a test would be valid only if a very accurate energy barrier associated with the process can be provided. Rate constants computed using Arrhenius-like expressions (as in both TST and NA-TST) are extremely sensitive to the employed energy barrier, where an error as small as 0.1 eV can result in orders of magnitude error in the corresponding rates. We are currently evaluating the possibility of performing very computationally intensive RASPT2 calculations on $[Ir(ppy)_2(bpy)]^+$ to obtain accurate energy barrier and non-radiative decay constant.

Now returning to the spin-mixed representation, one may expect that there are however CIs involving the two spin-mixed PESs, where the probability of NAE is very large. Nevertheless, assuming that the SOC between the $^3$MC and $S_0$ states is non-zero, no CIs between these spin-mixed states exist in the $S_0$ and $T_1$ two-state model.[31,43–45] To understand this, let us recall the condition for having a CI between two spin-pure diabatic states with the same spin. In such a framework (eq. S24-25 of Section S6) a CI is encountered when the two states have the same energy ($H_{el}^{00} = H_{el}^{11}$) and their coupling is zero ($H_{el}^{01} = 0$). In our case, the two spin-pure states $S_0$ and $^3$MC are our spin-diabatic states, while the corresponding spin-mixed states are the adiabatic states whose CI we are looking for. But while the first condition ($E(S_0) = E(\,^3MC)$) can be satisfied (as at the MECP), the condition $H_{el}^{01} = 0$ (i.e. $H_{SOC}^{S_0\,^3MC} = 0$ ) is never satisfied, because the $S_0$ and $^3$MC states are characterized by a large SOC.

Up to now we have ignored the zero-field splitting of the $T_1$ state into its three sublevels, i.e. $T_{-1,0,1}$.[46] In a recent work, Wang and Yarkony mathematically describe the transformation from spin-pure to spin-mixed states when considering a model with the $S_0$ state and the three sublevels composing the $T_1$ state.[31] They showed that two of the three sublevels have the same energy as the original $T_1$ state, whereas the third one increases its energy by a value equal to the SOC, and the spin-mixed $S_0$ state decreases its energy by a value equal to the SOC (see equations 7 of reference 31). We then conclude that when considering the three sublevels of $T_1$, the splitting at MECP between the spin-mixed state associated with $S_0$ and the lowest-energy components of the spin-mixed states associated with $T_1$, is equal to the SOC value. In our specific example, i.e. the $[Ir(ppy)_2(bpy)]^+$ complex, that corresponds to a 0.42 and 0.21 eV energy splitting at the $^3MC_{ax}/S_0$ and $^3MC_{eq}/S_0$ MECPs, respectively. These energies, although half the value obtained when ignoring the triplet zero-field splitting, are still very high, leading to small NAC (see eq 2).

The absence of the CIs does not mean that $^3$MC states are not involved in the non-radiative decay. $^3$MC states can still mediate a non-radiative decay in Ir3-TMCs, but, according to the presented model, not through a CI, so not in a way as efficient as in CI-mediated NAEs. When $^3$MC states are accessible and significantly lower in energy than any emitting state, they will play a role, as in the related Ru(II) $d^6$ complexes $[Ru(m\text{-}bpy)_3]^{2+}$ and $[Ru(tm\text{-}bpy)_3]^{2+}$, where their involvement was experimentally proven.[47] We propose that the relevance of a $^3$MC state is determined by how long Ir3-TMCs remain trapped on the $^3$MC PES (i.e. in the corresponding $^3$MC minimum). From the $^3$MC minimum, the decay back to $S_0$ through MECP will have low probability, as shown by our NA-TST calculation (see Section 5). However, if the excited state population remains trapped in the $^3$MC minimum, the small energy gap with the ground state will efficiently promote a $T_1$-to-$S_0$ ISC process, as in the emitting $T_1$ minima part of the population decay through a $T_1$-to-$S_0$ ISC process whose rate constant is normally indicated as $k_{ISC}$.[10,48] The described small $^3$MC-$S_0$ energy separation is caused by the broken coordination bond characterizing $^3$MC minima, in turn leading to a massive rise of the $S_0$ energy. In our example, i.e. $[Ir(ppy)_2(bpy)]^+$, the $^3$MC-$S_0$ gap at the $(^3MC_{ax})_{min}$ and $(^3MC_{eq})_{min}$ structures is 0.55 and 0.65 eV, respectively. A key difference is predictable between the $T_1$-to-$S_0$ ISC process operating at the $T_1$ emitting minima and at the $^3$MC equilibrium structures. In the former case, the $T_1$ and $S_0$ minima are normally nested states, and the process will follow the so-called energy gap law, so as the energy separation increases, the $T_1$-to-$S_0$ ISC transition probability decreases. In the latter case, the large geometrical difference between the $S_0$ and $^3$MC minima can instead lead to the opposite behavior, so as the energy separation increases, the $T_1$-to-$S_0$ ISC transition probability increases.[34,49] Further studies are required to describe the phenomena.

The scenario exemplified for $[Ir(ppy)_2(bpy)]^+$ is common to any Ir3-TMC, or at least to any Ir3-TMC in which $^3MC/S_0$ MECPs display similarly large SOC (1500 $cm^{-1}$), as it is in general the case (see Section S7). Sloped $^3MC/S_0$ MECPs are indeed present in any Ir3-TMC, as proven in Section S2. Actually, any $d^6$ TMC would describe such sloped $^3MC/S_0$ MECPs, so in case of presenting a large SOC between the involved $^3MC$ and $S_0$ states, again the same conclusions here exemplified for $[Ir(ppy)_2(bpy)]^+$ will be valid. This could be the case of Ru(II) TMCs, whose non-radiative decay mediated by $^3MC$ states is normally described in a similar way as for Ir(III) complexes,[50–54] and whose $^3MC/S_0$ MECPs usually display SOC values of the order of 1000 $cm^{-1}$.

In summary, we show that the role played by $^3MC$ states in the non-radiative deactivation of Ir3-TMCs deserves reconsideration. Specifically, we showed that it is incorrect to assume that at the $^3MC/S_0$ MECP the probability of decay back to the ground state minimum is equal to unity, and that, counterintuitively, the larger is the SOC between $^3MC$ and $S_0$ states, the lower is such probability. Moreover, we showed that, assuming a non-zero SOC between $^3MC$ and $S_0$ states for all geometries, the spin-mixed states resulting from the spin-pure $^3MC$ and $S_0$ states do not form CIs. Consequently, even if the established strategy of avoiding $^3MC$ population in order to minimize the non-radiative decay can still be valid, we postulate that the role of $^3MC$ states in the non-radiative decay of Ir3-TMCs is not directly related to the ability of reaching the $^3MC/S_0$ MECPs from the emitting minimum. Instead, we propose that the role of $^3MC$ states is dictated by the tendency of Ir3-TMCs to remain trapped in the $^3MC$ minima, from where, given sufficient time, the non-radiative decay to the ground state will occur by mean of a $T_1$-to-$S_0$ ISC process.

**Acknowledgements**

The financial support by the MCIN/AEI of Spain (projects PID2021-128569NB-I00 and CEX2024-001467-M, funded by MCIN/AEI/10.13039/501100011033 and by "ERDF A way of making Europe") and the Generalitat Valenciana (MFA/2022/017) is acknowledged. The MFA/2022/017 project is part of the Advanced Materials program supported by the MCIN with funding from the European Union NextGenerationEU (PRTR-C17.I1) and by Generalitat Valenciana. I.S.-D. also thanks Generalitat Valenciana for his predoctoral grant CIACIF/2021/438. A.G. acknowledges Grant RYC2023-044677-I, funded by MCIU/AEI/10.13039/501100011033 and FSE+. This material is partly based upon work supported by the U.S. Department of Energy, Office of Science, Office of Basic Energy Sciences Established Program to Stimulate Competitive Research under Award Number DE-SC0022178. This work was supported in part through the NYU IT High Performance Computing resources, services, and staff expertise."

Supporting Information
(Total of 14 pages)
for

# The role of spin-orbit coupling and state-crossing topography in the non-radiative decay of Ir(III) complexes

Iván Soriano-Díaz,[1] Ilya D. Dergachev,[2] Sergey A. Varganov,[3] Enrique Ortí,[1] and Angelo Giussani[1,*]

*[1] Institute for Molecular Science (ICMol), Universitat de València, Catedrático José Beltrán 2, 46980 Paterna, España.*

*[2] Department of Chemistry, University of Nevada, Reno, 1664 N. Virginia Street, Reno, NV 89557-0216, USA; Current address: Department of Chemistry, New York University, New York, New York 10003, USA*

*[3] Department of Chemistry, University of Nevada, Reno, 1664 N. Virginia Street, Reno, NV 89557-0216, USA*

Email: angelo.giussani@uv.es

**Section S1. Relevance of Ir(III) complexes in the field of electroluminescent devices and photodynamic therapy**

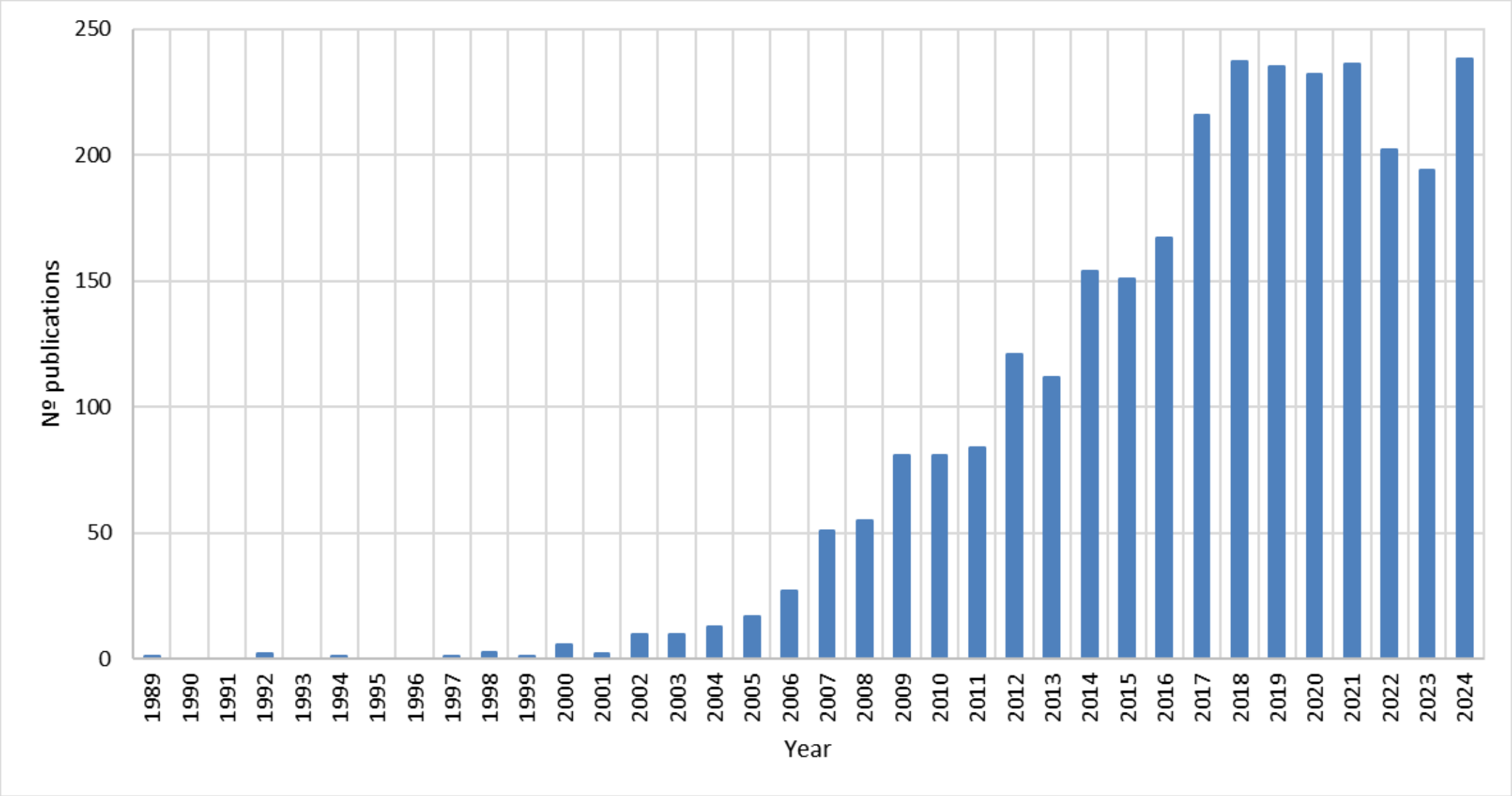

**Figure S1:** Number of publications per year on Web of Science studying electroluminescence applications (LECs or OLEDs) or photodynamic therapy (PDT) based on Ir(III) complexes. Search performed using TS=("Ir* III" OR "Ir(III)" OR "Ir (III)")) AND TS=("LECs" OR "LEECs" OR "electroluminescent application*" OR "photodynamic therapy" OR "OLEDs") as Booleans at 10/04/2025.

**Section S2. Triplet metal-centered states of Ir(III) and Ru(II) complexes**

The Ir(III) and Ru(II) metal centers possess electronic configurations of 5$d^6$ and 4$d^6$, respectively. In an octahedral coordination environment, the five *d* orbitals split into two groups according to ligand field theory: the lower-energy $t_{2g}$ ($d_{xy}$, $d_{xz}$, and $d_{yz}$) orbitals and the higher-energy $e_g$* ($d_{x^2-y^2}$ and $d_{z^2}$) orbitals. According to this theory, the $t_{2g}$ orbitals are stabilized by (2/5)$\Delta_o$ while the $e_g$* orbitals are destabilized by (3/5)$\Delta_o$ (see Figure S2a), where $\Delta_o$ represents the octahedral ligand field splitting energy.[1,2]

In principle, for a $d^6$ metal center, two electronic configurations are possible. In the low-spin configuration, all six electrons populate the $t_{2g}$ orbitals ($t_{2g}^6$), whereas in the high-spin configuration the electrons occupy both ($t_{2g}^4 e_g^{*2}$). The actual ground state is determined by the magnitude of $\Delta_o$ relative to the electron pairing energy. However, in Ir(III) and Ru(II) complexes this is high enough to favors low-spin configurations and leads to a unique singlet ground state, $S_0$. Eventhough, the choice of ligands, as reflected in the spectrochemical series, modify the magnitude $\Delta_o$.

According to molecular orbital theory, the $e_g$* orbitals from the metal center will combine with the $\sigma$ orbitals of the ligands confering them an antibonding character (see Figure S2b). Consequently, metal-centered ($^3$MC) excited states of Ir(III) and Ru(II) complexes imply an electron promotion into an $e_g$* orbital, which involves antibonding interactions between the metal center and the ligands. This causes that the $^3$MC minima display a dissociated coordination bond, which is instead present in both the ground state and the emitting minima, and in turn determines that, as explained in the main text, the $^3MC/S_0$ MECPs of Ir(III) and Ru(II) complexes have a sloped topology.

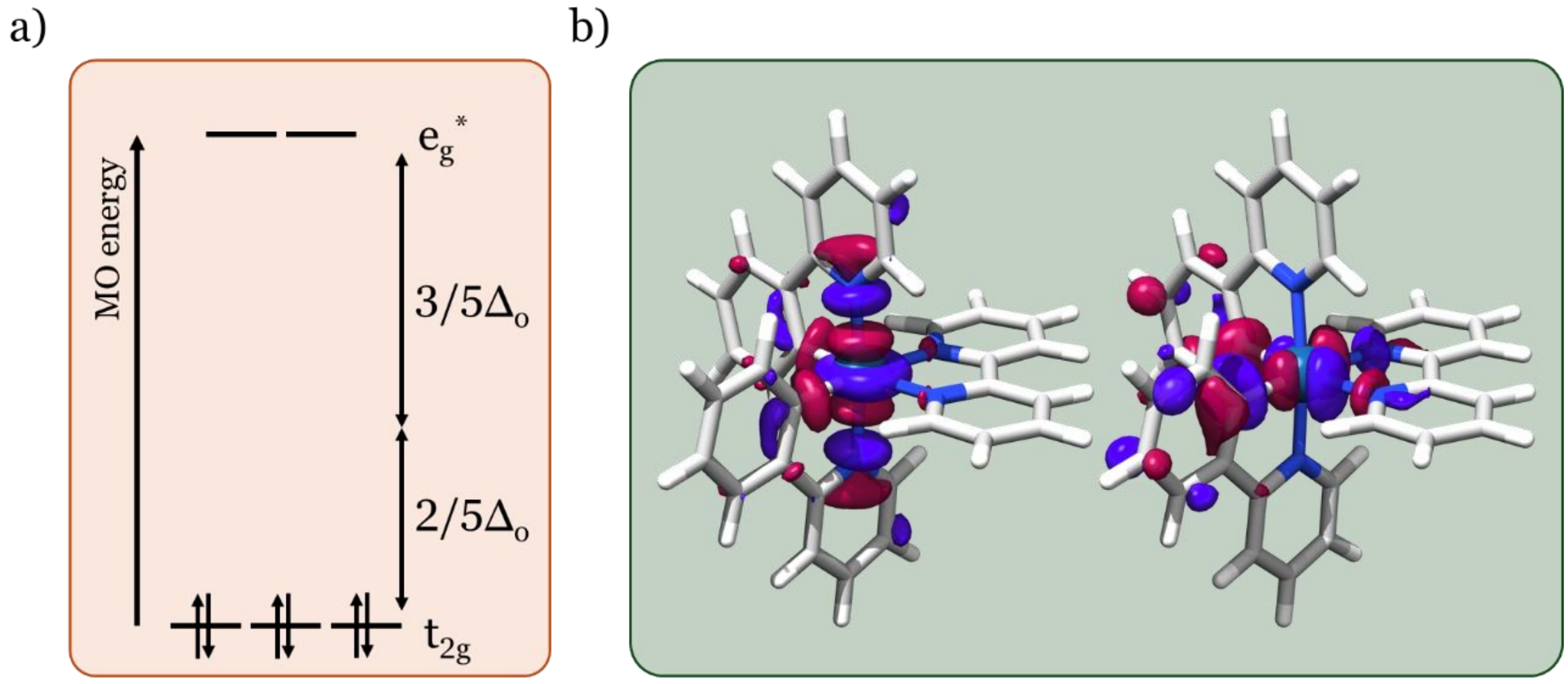

**Figure S2:** a) Schematic diagram showing the electronic configuration of $d^6$ octahedral metal complexes such as Ir(III) and Ru(II) complexes. b) $e_g$* antibonding molecular orbitals for $[Ir(ppy)_2bpy]^+$.

**Section S3. Derivation of the equations that govern NAEs working with spin-pure states and spin-mixed states.**

The time-dependent Schrödinger equation for a generic molecule can be written as follow:

$$i\dot{\Psi}(\boldsymbol{r},\boldsymbol{R},t) = \hat{H}_{mol}\Psi(\boldsymbol{r},\boldsymbol{R},t) \quad \text{eq. S1}$$

where $\hat{H}_{mol}$ is the molecular Hamiltonian and $\boldsymbol{r}, \boldsymbol{R}$ and $t$ correspond to the electronic, nuclear and time coordinates, respectively. $\hat{H}_{mol}$ is in turn composed of the nuclear kinetic operator $\hat{T}_R$, the electronic Hamiltonian $\hat{H}_{el}$, and the SOC term $\hat{H}_{SOC}$.

$$\hat{H}_{mol} = \hat{T}_R + \hat{H}_{el} + \hat{H}_{SOC} \quad eq. S2$$

The eigenfunctions of the electronic Hamiltonian $|s\rangle$ (i.e.: the adiabatic electronic states) form a complete basis set, so it is possible to expand the wave-function $\Psi(\boldsymbol{r},\boldsymbol{R},t)$ in such a basis set without introducing any approximation:

$$\hat{H}_{el}|s\rangle = V^s(\boldsymbol{R})|s\rangle \quad eq. S3$$

$$\Psi(\boldsymbol{r},\boldsymbol{R},t) = \sum_s \psi^s(\boldsymbol{R},t)\,|s\rangle \quad eq. S4$$

Substituting such an expression in the time-dependent Schrödinger equation S1, neglecting the $\hat{H}_{SOC}$ term, and projecting on the generic electronic state $|s'\rangle$, it is possible to obtain:

$$i\sum_s \dot{\psi}^s(\boldsymbol{R},t)\langle s'|s\rangle = \sum_s \langle s'|(\hat{T}_R + \hat{H}_{el})\psi^s(\boldsymbol{R},t)|s\rangle \quad eq. S5$$

It can be proved that:

$$\langle s'|\hat{H}_{el}\psi^s(\boldsymbol{R},t)|s\rangle = \delta_{ss'}V^{s'}(\boldsymbol{R})\psi^{s'}(\boldsymbol{R},t)|s'\rangle \quad eq. S6$$

$$\langle s'|\hat{T}_R\psi^s(\boldsymbol{R},t)|s\rangle = (\delta_{ss'}\hat{T}_R + \hat{\Lambda}^{ss'})\psi^s(\boldsymbol{R},t) \quad eq. S7$$

Where $V^{s'}(\boldsymbol{R})$ is the electronic energy of state $|s'\rangle$, and $\hat{\Lambda}^{ss'}$ are the so-called non-adiabatic couplings (NACs).[3,4]

Using these expressions, it is possible to arrive at the next equation that clearly shown that what causing that part of the population of the generic state $|s'\rangle$ to pass to another state $|s\rangle$, is the corresponding NAC ($\hat{\Lambda}^{ss'}$), which is in fact the term that in the next equation connects the time evolution of the population of state $|s'\rangle$ (i.e.: $\dot{\psi}^{s'}(\boldsymbol{R},t)$) with the population of the other state $|s\rangle$ (i.e.: $\psi^{s}(\boldsymbol{R},t)$):

$$i\dot{\psi}^{s'}(\boldsymbol{R},t) = \left(\hat{T}_R + V^{s'}(\boldsymbol{R})\right)\psi^{s'}(\boldsymbol{R},t) + \sum_{s} \hat{\Lambda}^{ss'}\,\psi^{s}(\boldsymbol{R},t) \quad eq.S8$$

The NACs are mostly composed of the so-called first-order non-adiabatic coupling, which, using the Hellmann-Feynman theorem can be proven to be inversely proportional to the energy separation of the two involved states:

$$\hat{\Lambda}^{ss'} \cong \langle s|\hat{\nabla}_R|s'\rangle = \frac{\langle s|\hat{\nabla}_R \hat{H}_{el}|s'\rangle}{V^{s'}(\boldsymbol{R}) - V^{s}(\boldsymbol{R})} \quad eq.S9$$

In the particular case in which we expand the wave-function in the basis of only two adiabatic electronic states of pure-spin having the same spin symmetry, and for example the $S_0$ and $S_1$ states, we obtain, considering that the initial population is in the $S_1$ state:

$$i\dot{\psi}^{S_1}(\boldsymbol{R},t) = \left(\hat{T}_R + V^{S_1}(\boldsymbol{R})\right)\psi^{S_1}(\boldsymbol{R},t) + \hat{\Lambda}^{S_0S_1}\psi^{S_0}(\boldsymbol{R},t) \quad eq.S10$$

And the corresponding IC process will have a higher probability at regions of CI between the $S_0$ and $S_1$ states.

Let's again write the wave function on the basis of adiabatic electronic states of pure-spin, but now allowing them to have different spin symmetry. Considering the $\hat{H}_{SOC}$ term in the molecular Hamiltonian appearing in the time-dependent Schrödinger equation,[5] and working in an analogous way as before, we obtain:

$$\langle s'|\hat{H}_{el}\psi^{s}(\boldsymbol{R},t)|s\rangle = \delta_{ss'}V^{s'}(\boldsymbol{R})\psi^{s'}(\boldsymbol{R},t)|s'\rangle \quad eq.S11$$

$$\langle s'|\hat{T}_R\psi^{s}(\boldsymbol{R},t)|s\rangle = (\delta_{ss'}\hat{T}_R + \delta_{\sigma(s)\sigma(s')}\hat{\Lambda}^{ss'})\psi^{s}(\boldsymbol{R},t) \quad eq.S12$$

$$\langle s'|\hat{H}_{SOC}\psi^{s}(\boldsymbol{R},t)|s\rangle = (1 - \delta_{\sigma(s)\sigma(s')})H_{SOC}^{s's}\psi^{s}(\boldsymbol{R},t) \quad eq.S13$$

Notice that in eq.S12 the NAC term appears only for state of the same spin symmetry (ensure by the presence of the $\delta_{\sigma(s)\sigma(s')}$ term, where the symbol $\sigma(s)$ indicate the spin coordinate of state $|s\rangle$), while in eq. S13 the SOC term appears only for state of different spin symmetry (ensure by the presence of the $1-\delta_{\sigma(s)\sigma(s')}$ term).

Now using these expressions and considering the specific case of just one singlet and triplet state (i.e.: the $S_0$ and $T_1$ states) we can write:

$$i\dot{\psi}^{T_1}(\boldsymbol{R},t) = \left(\hat{T}_R + V^{T_1}(\boldsymbol{R})\right)\psi^{T_1}(\boldsymbol{R},t) + H_{SOC}^{S_0T_1}\psi^{S_0}(\boldsymbol{R},t) \quad eq.S14$$

So now what connects the time evolution of the population of initially populated triplet state (i.e.: $\dot{\psi}^{T_1}(\boldsymbol{R},t)$) with the population of the ground state (i.e.: $\psi^{S_0}(\boldsymbol{R},t)$), is the $H_{SOC}^{S_0T_1}$ term.

Finally, we will use the spin-mixed states $|SM\rangle$ resulting from the diagonalization of the Hamiltonian formed by the electronic Hamiltonian plus the $\hat{H}_{SOC}$ term in order to expand the wave-function, again in the case of considering just two states:

$$\begin{pmatrix} V^{S_0} & H_{SOC}^{S_0T_1} \\ H_{SOC}^{S_0T_1} & V^{T_1} \end{pmatrix} \xrightarrow{diagonalization} \begin{pmatrix} V^{SM_0} & 0 \\ 0 & V^{SM_1} \end{pmatrix} \quad eq.S15$$

$$|SM_1\rangle = c_{S_0}^{SM_1}|S_0\rangle + c_{T_1}^{SM_1}|T_1\rangle \quad eq.S16$$

$$|SM_2\rangle = c_{S_0}^{SM_2}|S_0\rangle + c_{T_1}^{SM_2}|T_1\rangle \quad eq.S17$$

Proceeding as before

$$\langle SM'|(\hat{H}_{el} + \hat{H}_{SOC})\psi^{SM}(\boldsymbol{R},t)|SM\rangle = \delta_{SMSM'}V^{SM'}(\boldsymbol{R})\psi^{SM'}(\boldsymbol{R},t)|SM'\rangle \quad eq.S18$$

$$\langle SM'|\hat{T}_R\psi^{SM}(\boldsymbol{R},t)|SM\rangle = (\delta_{SMSM'}\hat{T}_R + \hat{\Lambda}^{SMSM'})\psi^{SM}(\boldsymbol{R},t) \quad eq.S19$$

That for the specific case of just two spin-mixed states resulting from the combination of the original $S_0$ and $T_1$ states, leads to the next equation:

$$i\dot{\psi}^{SM_1}(\boldsymbol{R},t) = \left(\hat{T}_R + V^{SM_1}(\boldsymbol{R})\right)\psi^{SM_1}(\boldsymbol{R},t) + \hat{\Lambda}^{SM_0SM_1}\psi^{SM_0}(\boldsymbol{R},t) \quad eq.S20$$

Eq. S20 clearly shows that what couple spin-mixed states then allowing for non-adiabatic events to occurs, are again the NACs, whose value again is inversely proportional to the energy separation of the two involved spin-mixed states.

## Section S4. Computational details used for calculating the SOC values of $[Ir(ppy)_2(bpy)]^+$.

The values of the SOC have been obtained by performing SOC-TDDFT calculations with ORCA 5.04,[6] using the PBE0 functional,[7] employing the ZORA Hamiltonian[8] to simulate the relativistic effects, a mean-field spin–orbit operator,[9] and the basis-set ZORA-def2-SVP[10] for non-Ir atoms and ZORA-def2-TZVP[11] for the Ir center.

## Section S5. Calculating non-radiative decay probabilities using the NAST code.

In this work, we consider a sloped intersection of two spin-diabatic states with different spin multiplicity strongly coupled by the spin-orbit coupling at MECP. These states are shown in Figure S3 (see also Figure 2b in the main text). The relaxation process starts at the minimum of the spin-diabatic potential of the excited $^3$MC state (solid blue line) and propagates towards the MECP with the ground $S_0$ state potential (solid orange line). Large spin-orbit coupling (>1000 $cm^{-1}$) between the two states at MECP results in the population transfer to the upper branch of the spin-diabatic potential of the ground state. When the system runs out of kinetic energy and the population starts moving in the opposite direction, the system will pass through MECP a second time and a second population transfer back to the potential of the excited triplet state occurs. Thus, the probability of the population transfer to the lower branch of the spin-diabatic potential of the ground state becomes very small. This means that the probability of re-population of the ground state Franck-Condon region becomes very small. In the adiabatic representation, a large spin-orbit coupling results in a large energy gap between the two adiabatic potentials. Thus, a sloped intersection of two states of different spin multiplicities and their large coupling via spin-orbit interaction results in the population locked on the upper adiabatic potential (dashed red line), making the transition probability to the lower spin-adiabatic state (dashed green line) very small.

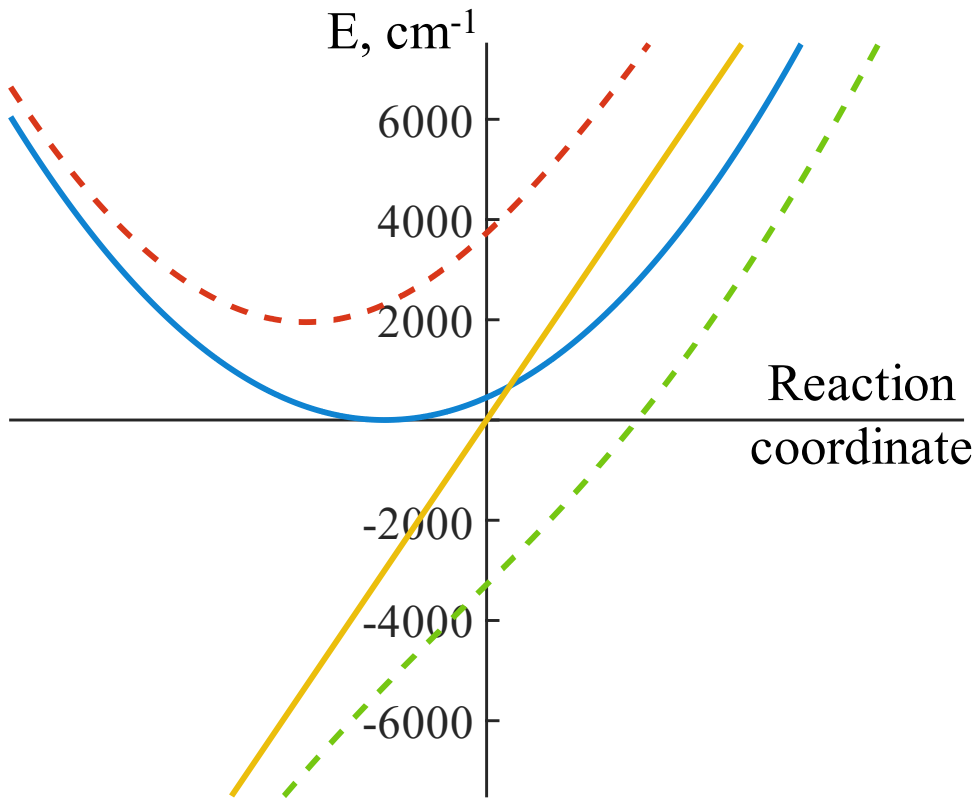

**Figure S3.** Sloped intersection of two spin-diabatic potentials (solid lines) and two spin-adiabatic potentials (dashed lines) after adiabatization with spin-orbit coupling.

In order to quantitatively support the conclusion that sloped MECPs characterized by large SOC do not efficiently mediate the re-population of the ground state Franck-Condon region, we calculate such a probability for the specific case of the path from the $^3MC_{eq}$ minimum to the $^3MC_{eq}/S_0$ MECP described in the main text for the $[Ir(ppy)_2(bpy)]^+$ using the NAST code.

At the MECP defined by two spin-diabatic PESs, the probability of transition between the corresponding spin-adiabatic surfaces can be computed according to the following Landau-Zener equation:

$$p_{\mathrm{LZ}}(\varepsilon_\perp) = \exp\left(-\frac{2\pi H_{\mathrm{SO}}^2}{\hbar|\Delta\mathbf{g}|}\sqrt{\frac{\mu_\perp}{2(\varepsilon_\perp - E_X)}}\right) \quad eq.S21$$

where $H_{\mathrm{SO}}$ is the spin-orbit coupling of the two spin-diabatic PESs at MECP, $\hbar$ is the reduced Planck's constant, $|\Delta\mathbf{g}|$ is the norm of the gradient parallel to the reaction coordinate at MECP, $\mu_\perp$ is the reduced mass along the reaction coordinate at MECP, $\varepsilon_\perp$ is the reaction coordinate energy (i.e. the energy of the system along the coordinate leading to MECP, which in this specific case is an Ir-N coordination bond), and $E_{\mathrm{X}}$ is the MECP energy barrier with respect to the reactant minimum (in this case the $^3MC_{eq}$ minimum).
Considering the case above described, according to which the system can pass two times through MECP, the probability of decaying to the ground state Franck-Condon region (i.e. the probability of population transfer to the lower branch of the spin-diabatic potential of the ground state) can be computed with the following equation:[12]

$$P_{LZ}(\varepsilon_\perp) = 2 * [p_{\mathrm{LZ}}(\varepsilon_\perp) - p_{\mathrm{LZ}}(\varepsilon_\perp)^2] \quad eq.S22$$

Figure S4 present a plot of the probability of decaying to the ground state Franck-Condon region at MECP as a function of the reaction coordinate energy ($\varepsilon_\perp$) for the path defined by the $^3MC_{eq}$ minimum and $^3MC_{eq}/S_0$ MECP of $[Ir(ppy)_2(bpy)]^+$. It is possible to appreciate that even at the very high value of the reaction coordinate energy equal to 1 eV, corresponding to the case of having 1 eV in the coordinate connecting the $^3MC_{eq}$ and $^3MC_{eq}/S_0$ MECP structures, the probability of decay is still significantly lower than 1 (specifically, it is equal to 0.24).

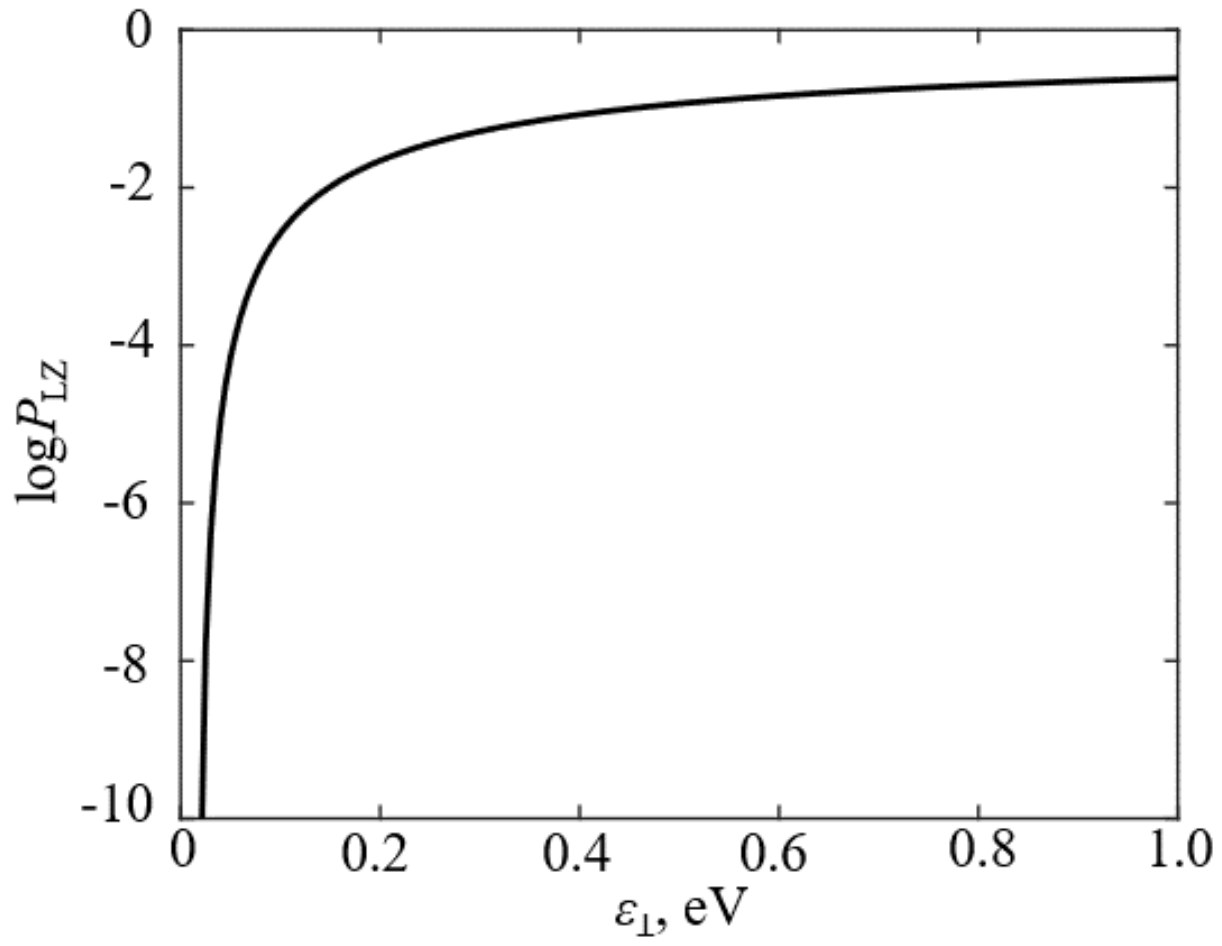

**Figure S4.** Probability of decaying to the ground state Franck-Condon region at MECP, computed using equation S22, as a function of the reaction coordinate energy ($\varepsilon_\perp$) for the path defined by the $^3MC_{eq}$ minimum and $^3MC_{eq}/S_0$ MECP of $[Ir(ppy)_2(bpy)]^+$.

**Section S6. Required conditions for having a CI in the framework of adiabatic and diabatic states.**

Considering two adiabatic states $a^0$ and $a^1$ of pure-spin having the same spin symmetry, by definition a CI is a point in which the two states have the same energy. In terms of the matrix elements of the electronic Hamiltonian ($H_{el}^{ij}$), and remembering that adiabatic states are by definition eigenfunctions of the electronic Hamiltonian, we have a CI when the following condition is fulfilled.

$$H_{el}^{a^0a^0} = H_{el}^{a^1a^1} \quad eq. S23$$

Considering instead two diabatic states $d^0$ and $d^1$ derivable form a unitary transformation of the $a^0$ and $a^1$ adiabatic states, and remembering that diabatic states are no longer eigenfunction of the electronic Hamiltonian, it is possible to prove that we have a CI when the following two conditions are fulfilled:[13]

$$H_{el}^{d^0d^0} = H_{el}^{d^1d^1} \quad eq. S24$$

$$H_{el}^{d^0d^1} = 0 \quad eq. S25$$

Two spin-pure adiabatic states of different spin symmetry (as for example a singlet state and a triplet state), obtained as eigenfunction of the spin-free electronic Hamiltonian, behave as diabatic states and in fact the corresponding PES can cross. In such a case equation 22 can be fulfilled, while, as long as there is SOC between the two, equation 23 will never be satisfied. Consequently, as long there is a non-null SOC between two spin-pure states of different spin symmetry, there will never be a CI for the corresponding spin-mixed states.

**Section 7. Examples of Ir(III) and Ru(II) complexes whose $^3$MC minima display a broken coordination bond and a large SOC with the ground state.**

## Ir(III) Complexes

### $[Ir(ppy)_2bpy]^+$ ref. 14-15

reMC$_{ax\,1}$

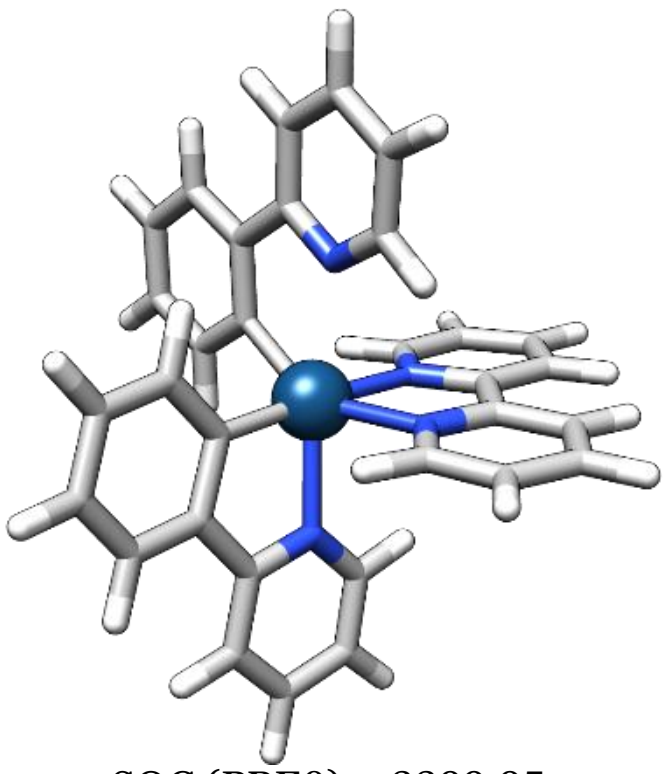

SOC (PBE0) = 3399.95
SOC (B3LYP) = 3362.43

MC$_{ax\,2}$

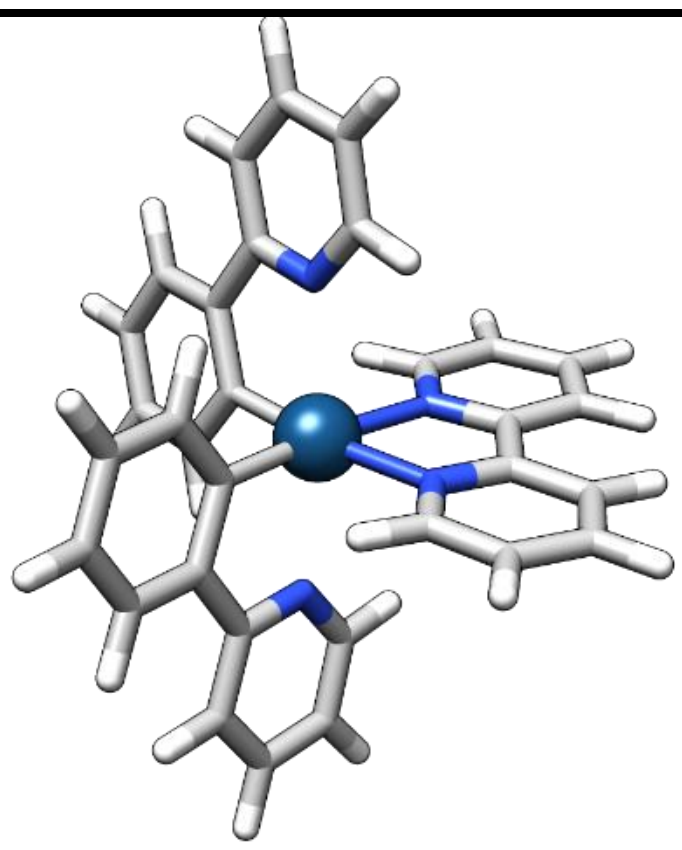

SOC (PBE0) = 3044.23
SOC (B3LYP) = 1683.64

MC$_{eq\,1}$

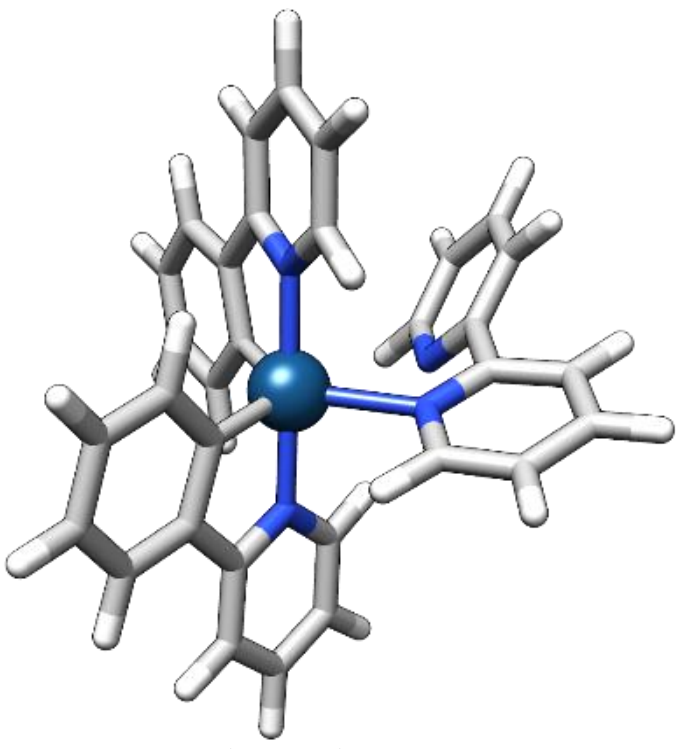

SOC (PBE0) = 1676.19
SOC (B3LYP) = 1897.39

### $[Ir(ppy)_2bpyph]^+$ ref. 14-15

MC$_{ax\,1}$

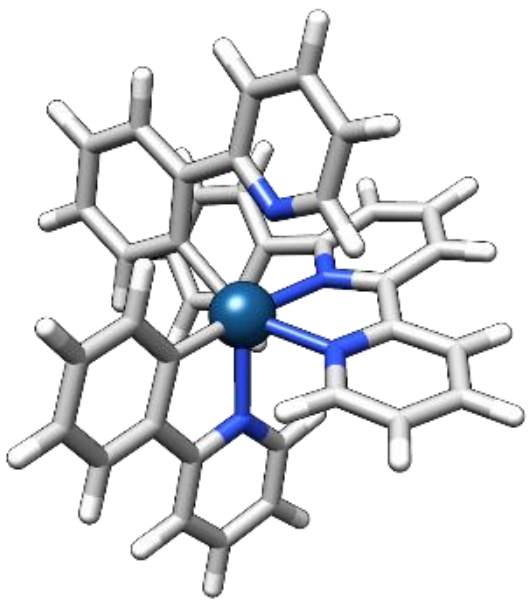

SOC (PBE0) = 3445.39
SOC (B3LYP) = 3471.38

MC$_{ax\,2}$

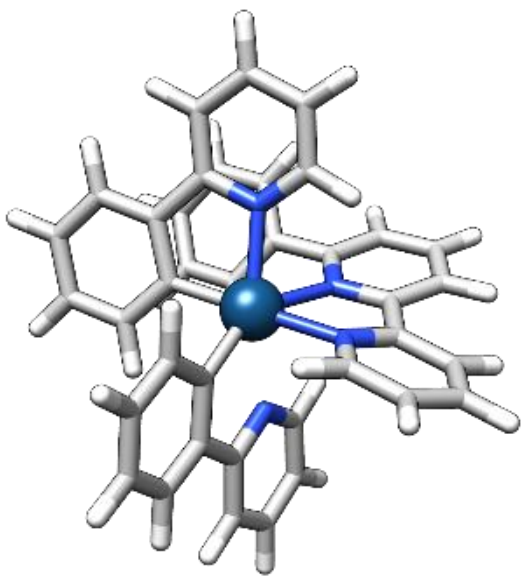

SOC (PBE0) = 3456.22
SOC (B3LYP) = 3467.86

MC$_{eq\,1}$

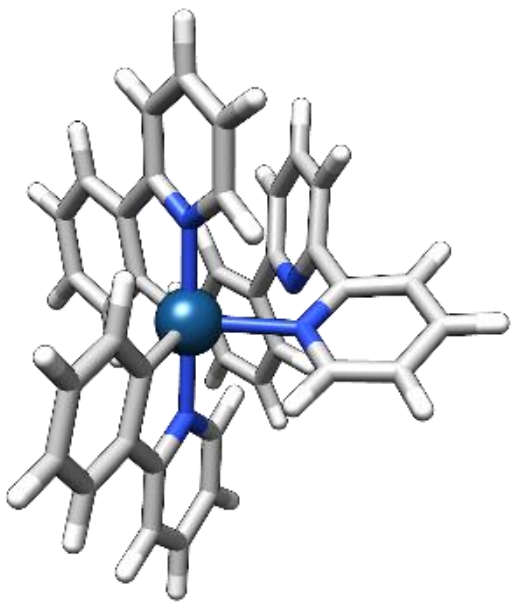

SOC (PBE0) = 1230.39
SOC (B3LYP) = 1505.32

MC$_{eq\,2}$

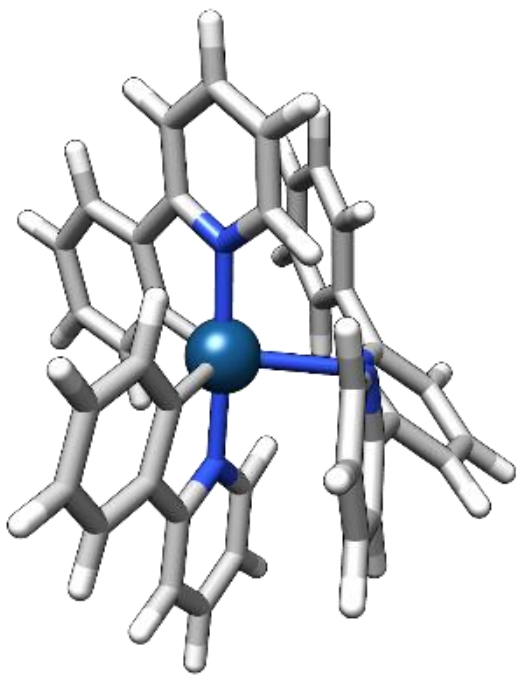

SOC (PBE0) = 1747.99
SOC (B3LYP) = 2028.45

### $[Ir(ppy)_2bpyph_2]^+$ ref. 14-15

MC$_{ax\,1}$

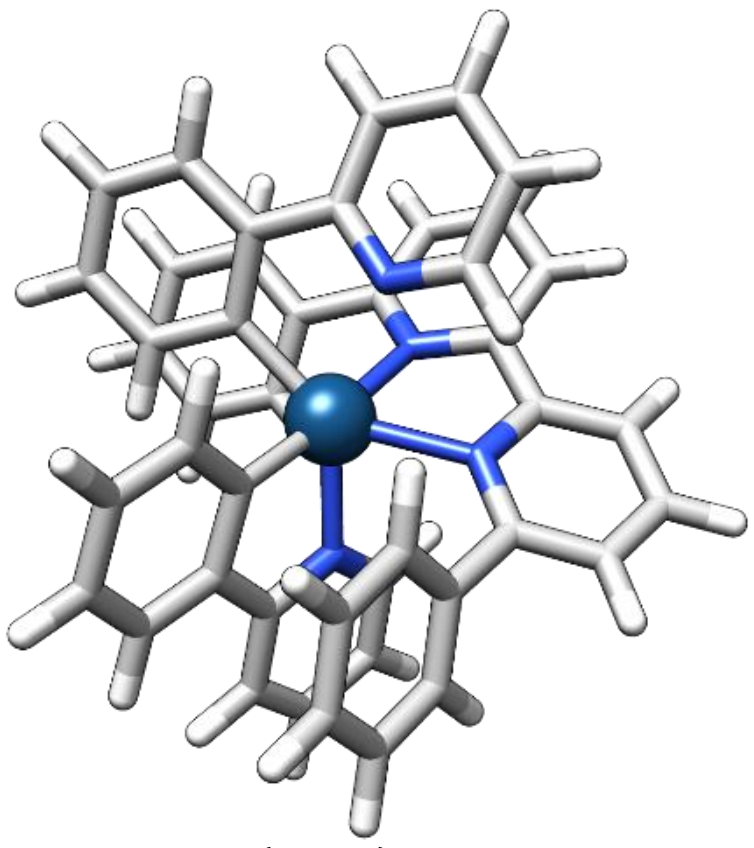

SOC (PBE0) = 2978.27
SOC (B3LYP) = 3165.88

MC$_{eq\,1}$

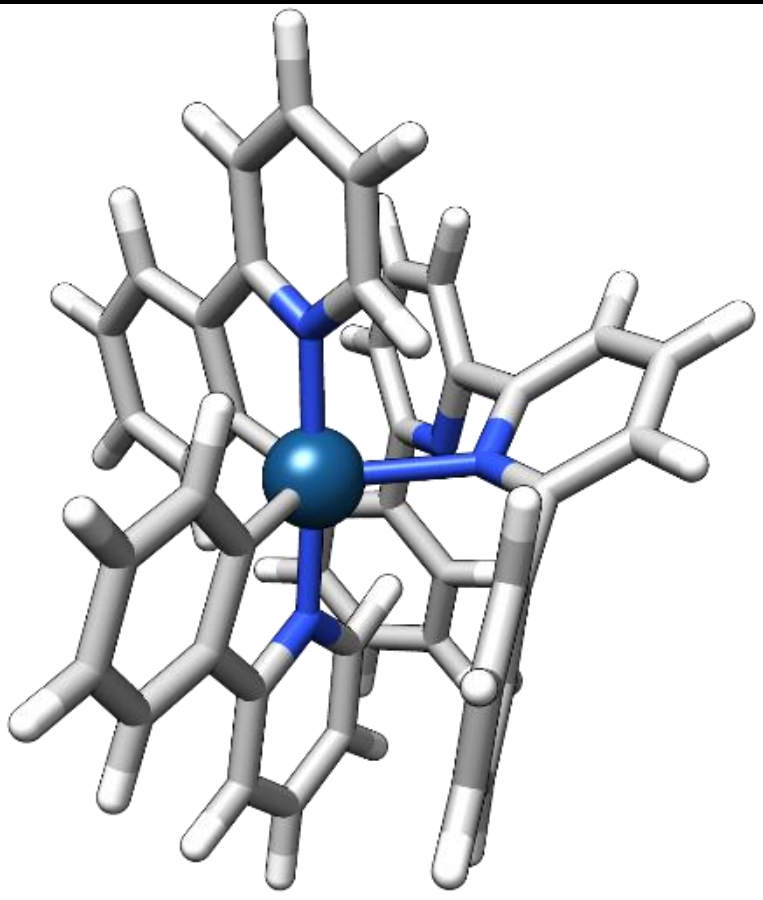

SOC (PBE0) = 1755.85
SOC (B3LYP) = 2037.63

| [Ir(diF-ppy)$_2$dtb-bpy]$^{+}$ ref. 16 | | |
|---|---|---|
| MC$_{ax\,1}$ | MC$_{ax\,2}$ | MC$_{eq\,1}$ |
| 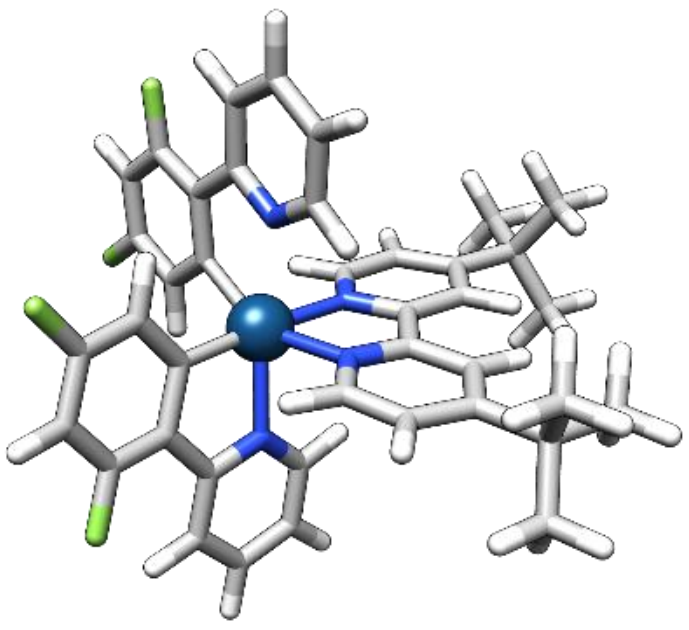<br>SOC (PBE0) = 3502.76<br>SOC (B3LYP) = 3507.02 | 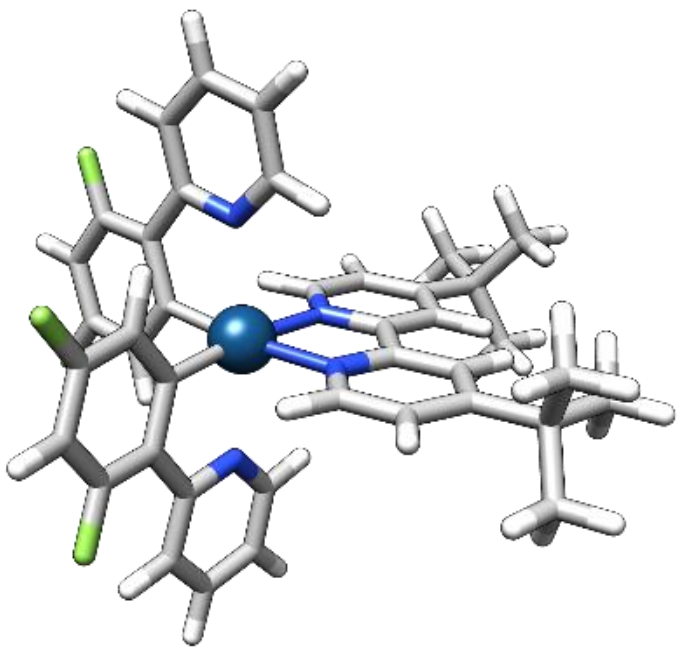<br>SOC (PBE0) = 3442.92<br>SOC (B3LYP) = 3102.19 | 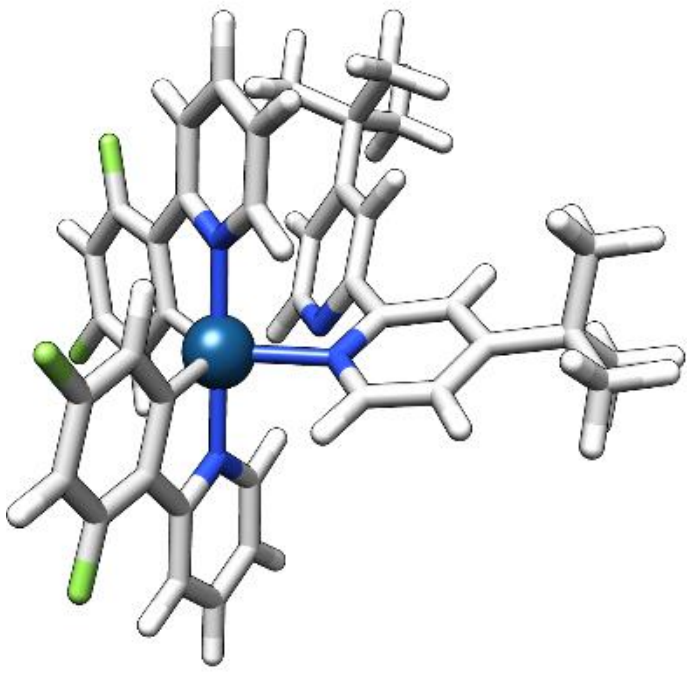<br>SOC (PBE0) = 1231.65<br>SOC (B3LYP) = 1826.56 |

| [Ir(ppy)$_2$(pyim)]$^{+}$ ref. 16-17 | | | |
|---|---|---|---|
| MC$_{ax\,1}$ | MC$_{ax\,2}$ | MC$_{eq\,1}$ | MC$_{eq\,2}$ |
| 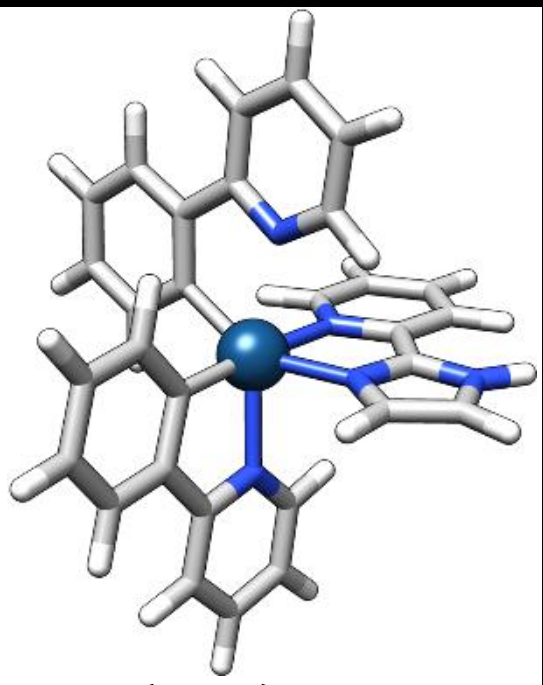<br>SOC (PBE0) = 1224.23<br>SOC (B3LYP) = 3559.18 | 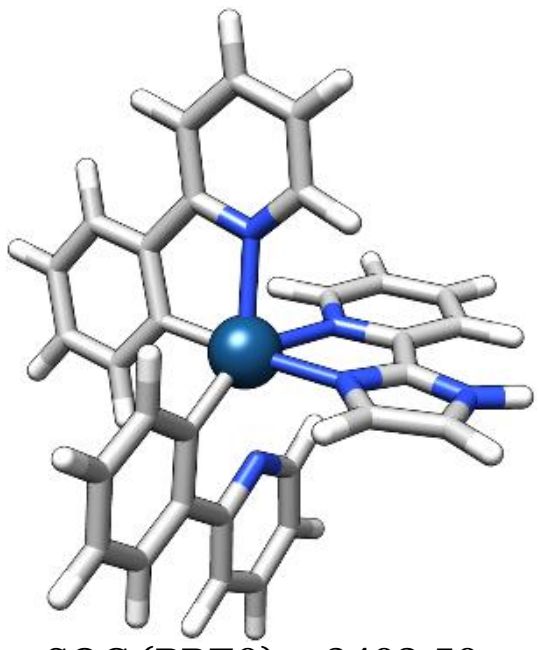<br>SOC (PBE0) = 3402.50<br>SOC (B3LYP) = 3362.21 | 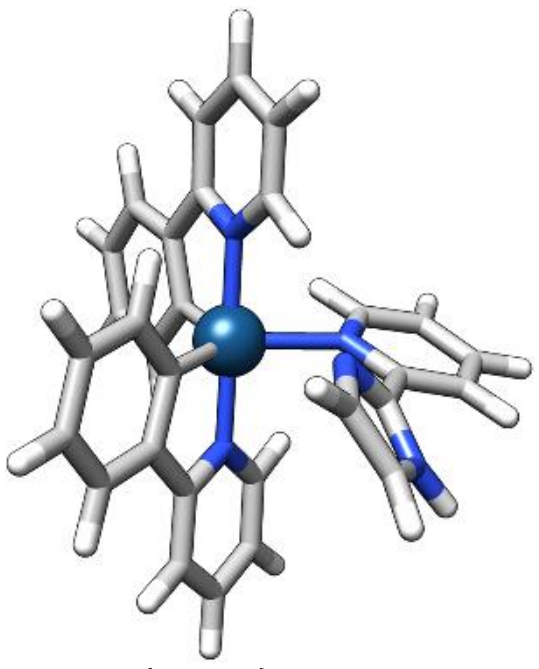<br>SOC (PBE0) = 1733.03<br>SOC (B3LYP) = 1939.37 | 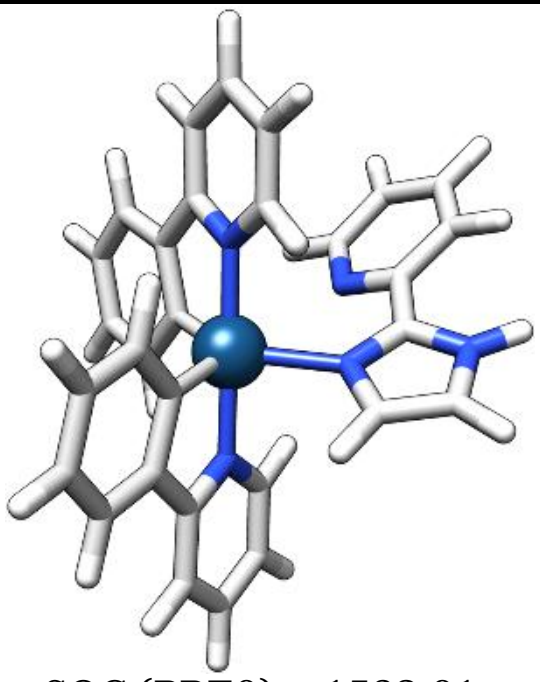<br>SOC (PBE0) = 1532.01<br>SOC (B3LYP) = 1754.63 |

| [Ir(ppy)$_2$(pyMebim)]$^{+}$ ref. 17 | | | |
|---|---|---|---|
| MC$_{ax\,1}$ | MC$_{ax\,2}$ | MC$_{eq\,1}$ | MC$_{eq\,2}$ |
| 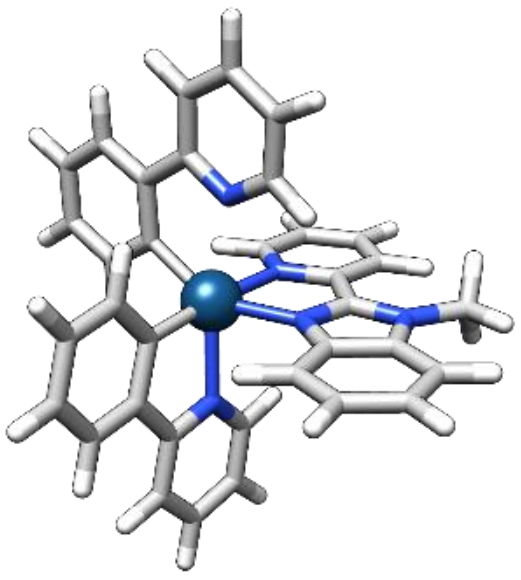<br>SOC (PBE0) = 3512.28<br>SOC (B3LYP) = 3484.94 | 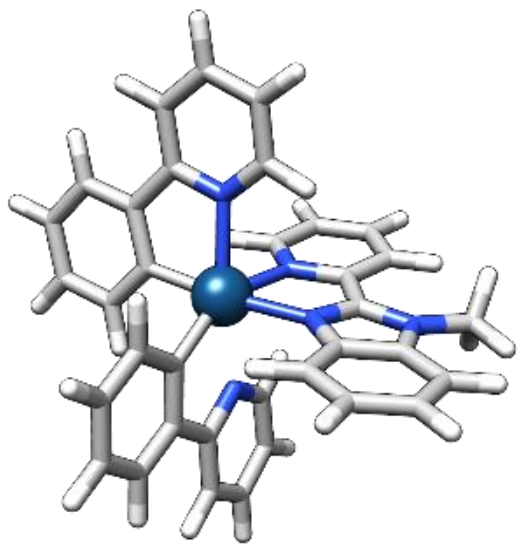<br>SOC (PBE0) = 3497.86<br>SOC (B3LYP) = 3502.53 | 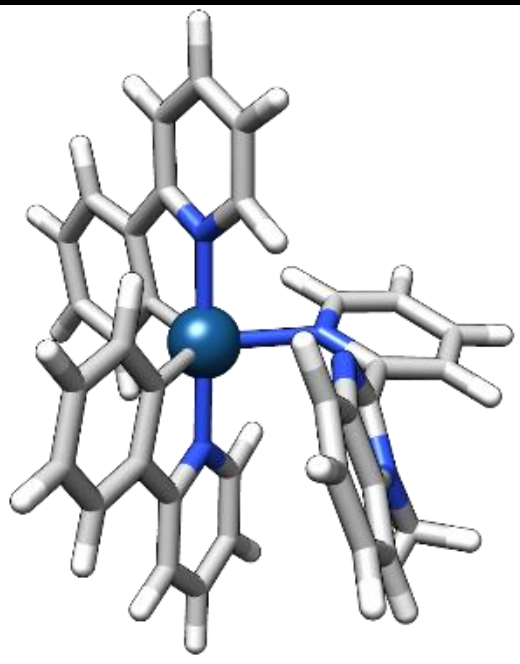<br>SOC (PBE0) = 1311.46<br>SOC (B3LYP) = 1562.65 | 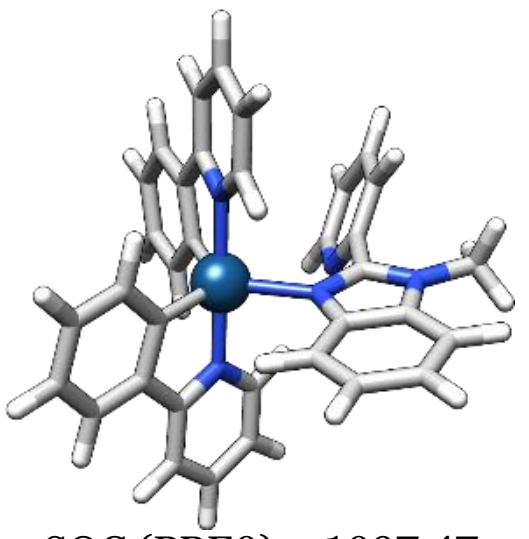<br>SOC (PBE0) = 1097.47<br>SOC (B3LYP) = 1345.07 |

| [Ir(ppy)$_2$(PIT)]$^{+}$ ref. 18 | | | |
|---|---|---|---|
| MC$_{ax\,1}$ | MC$_{ax\,2}$ | MC$_{eq\,1}$ | MC$_{eq\,2}$ |
| 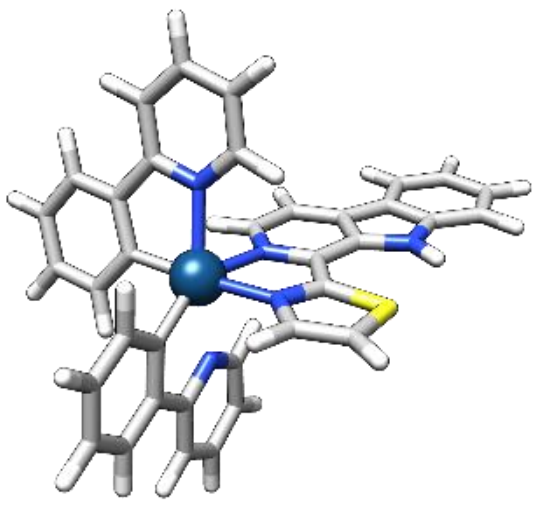<br>SOC (PBE0) = 3385.49<br>SOC (B3LYP) = 3360.87 | 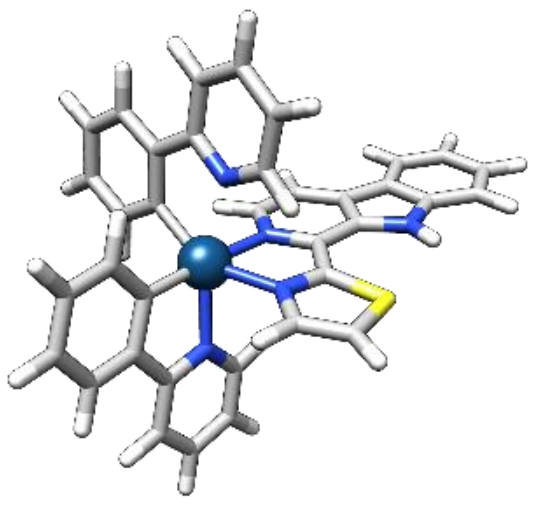<br>SOC (PBE0) = 3481.33<br>SOC (B3LYP) = 3460.07 | 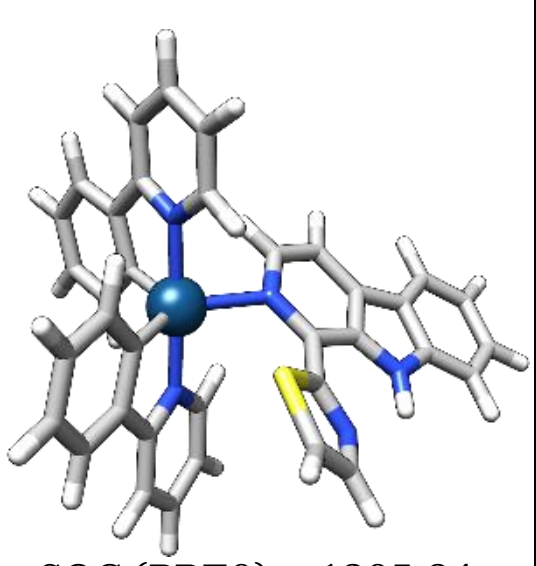<br>SOC (PBE0) = 1205.84<br>SOC (B3LYP) = 1900.99 | 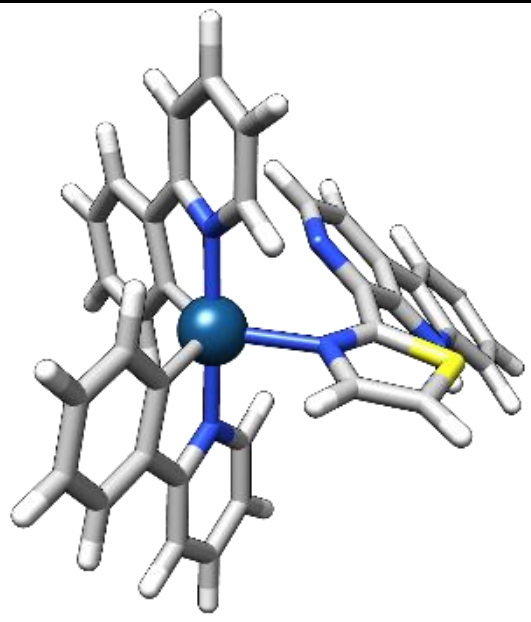<br>SOC (PBE0) = 1364.10<br>SOC (B3LYP) = 2023.83 |

| $[Ir(ppy)_2(MePIT)]^{+}$ ref. 18 | | | |
|---|---|---|---|
| $MC_{ax\,1}$ | $MC_{ax\,2}$ | $MC_{eq\,1}$ | $MC_{eq\,2}$ |
| 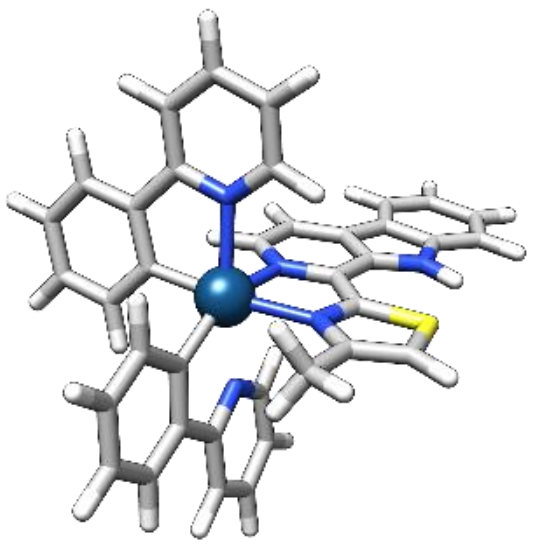<br>SOC (PBE0) = 3484.68<br>SOC (B3LYP) = 3508.76 | 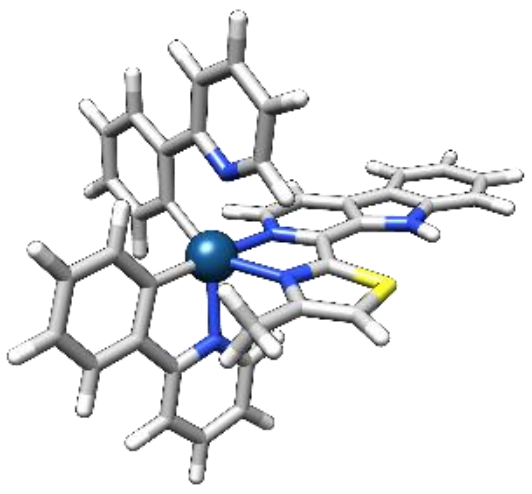<br>SOC (PBE0) = 3453.01<br>SOC (B3LYP) = 3453.19 | 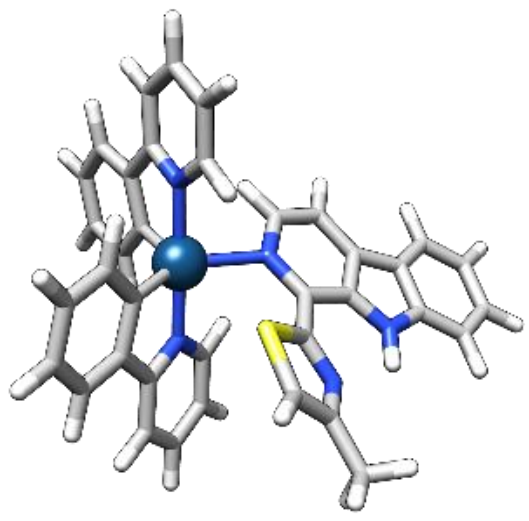<br>SOC (PBE0) = 1626.13<br>SOC (B3LYP) = 1904.12 | 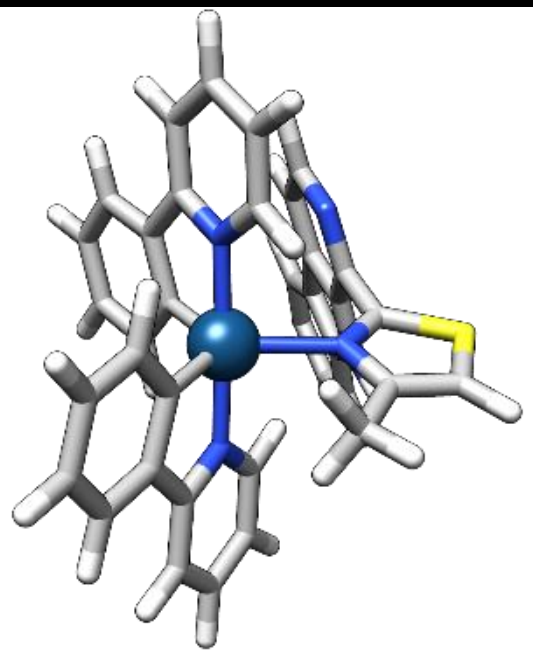<br>SOC (PBE0) = 1226.82<br>SOC (B3LYP) = 1500.36 |
| $[Ir(ppy)_2(PhPIT)]^{+}$ ref. 18 | | | |
| $MC_{ax\,1}$ | $MC_{ax\,2}$ | $MC_{eq\,1}$ | $MC_{eq\,2}$ |
| 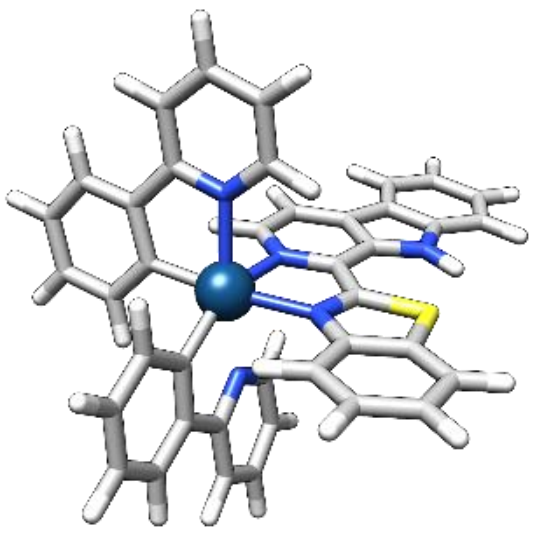<br>SOC (PBE0) = 3478.68<br>SOC (B3LYP) = 3497.71 | 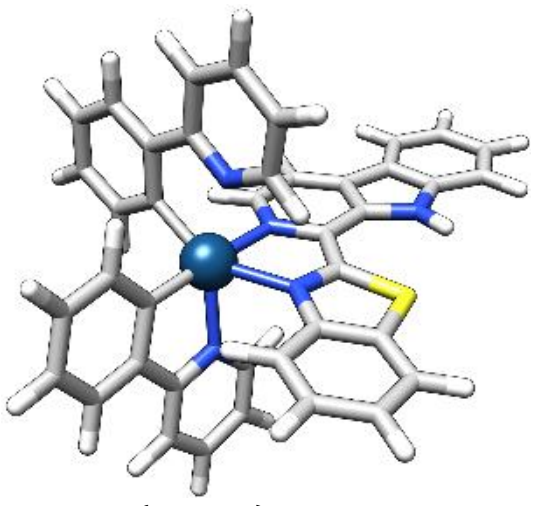<br>SOC (PBE0) = 3426.45<br>SOC (B3LYP) = 3427.13 | 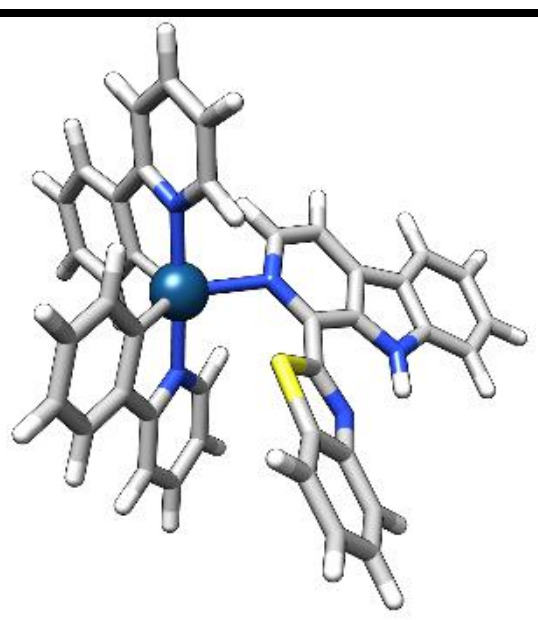<br>SOC (PBE0) = 1643.32<br>SOC (B3LYP) = 1933.49 | 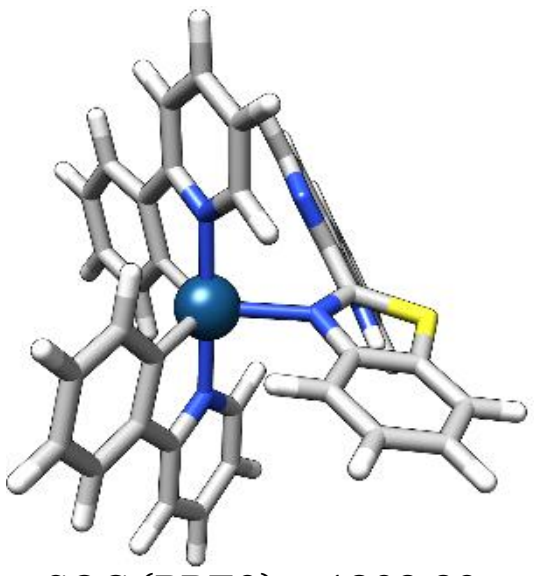<br>SOC (PBE0) = 1293.39<br>SOC (B3LYP) = 1855.36 |

**Figure S5:** Optimized geometries calculated for the $^3$MC minima (axial and equatorial) of a series of Ir(III) complexes. The value of the SOC (in $cm^{-1}$) between the $^3$MC and $S_0$ states computed at the SOC-TDDFT PBE0/(ZORA-def2-TZVP + ZORA-def2-SVP level, SOC(PBE0), and at the SOC-TDDFT B3LYP/(ZORA-def2-TZVP + ZORA-def2-SVP level, SOC(B3LYP), is also reported.

| Ru(II) complexes | | |
|---|---|---|
| [Ru(bpy)$_2$(PIT)]$^{+2}$ ref. 17 | | |
| MC$_{ax\ 1}$ | MC$_{ax\ 2}$ | MC$_{ax\ 3}$ |
| 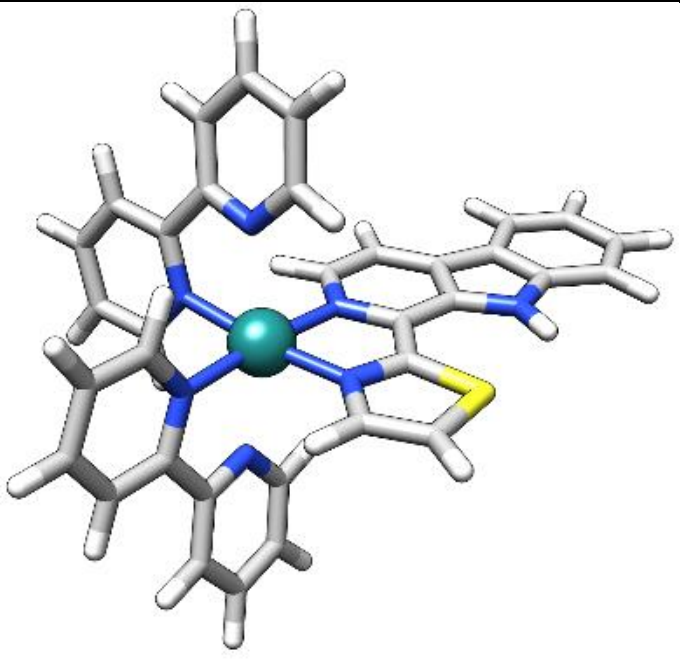<br>SOC (PBE0) = 1043.75<br>SOC (B3LYP) = 1042.01 | 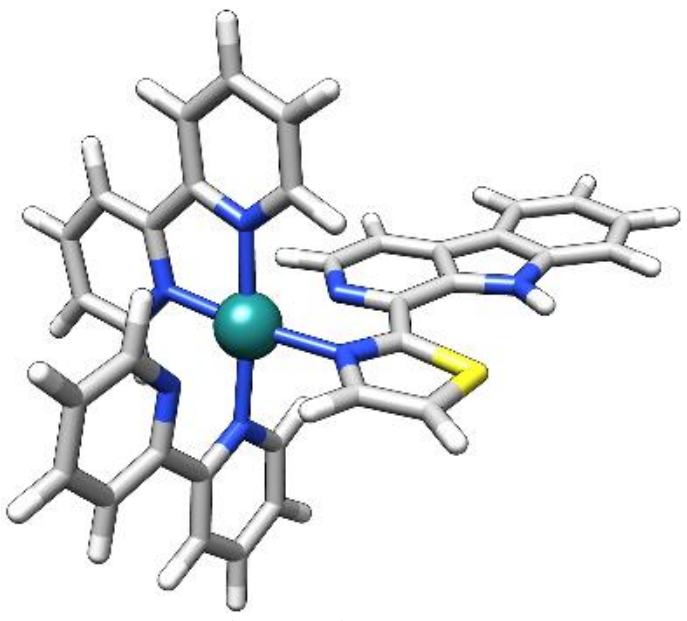<br>SOC (PBE0) = 1030.54<br>SOC (B3LYP) = 1023.59 | 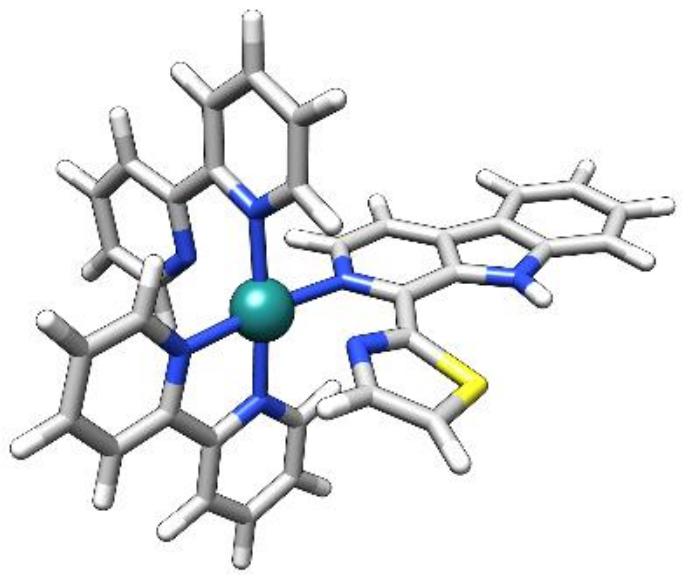<br>SOC (PBE0) = 1030.08<br>SOC (B3LYP) = 1024.18 |
| [Ru(bpy)$_2$(MePIT)]$^{+2}$ ref. 17 | | |
| MC$_{ax\ 1}$ | MC$_{ax\ 2}$ | MC$_{ax\ 3}$ |
| 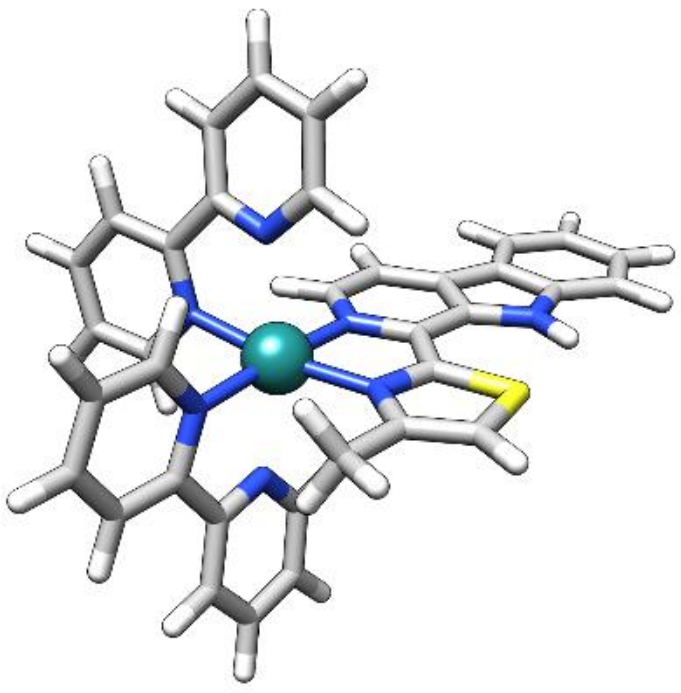<br>SOC (PBE0) = 1055.20<br>SOC (B3LYP) = 1053.75 | 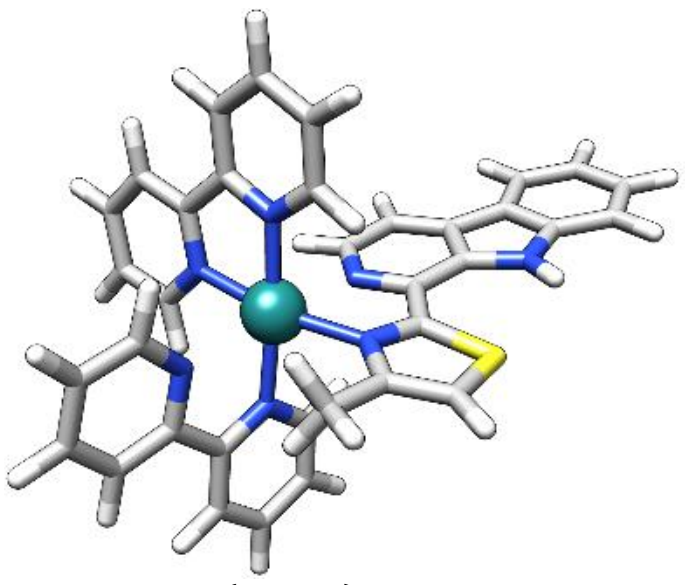<br>SOC (PBE0) = 1034.64<br>SOC (B3LYP) = 1034.36 | 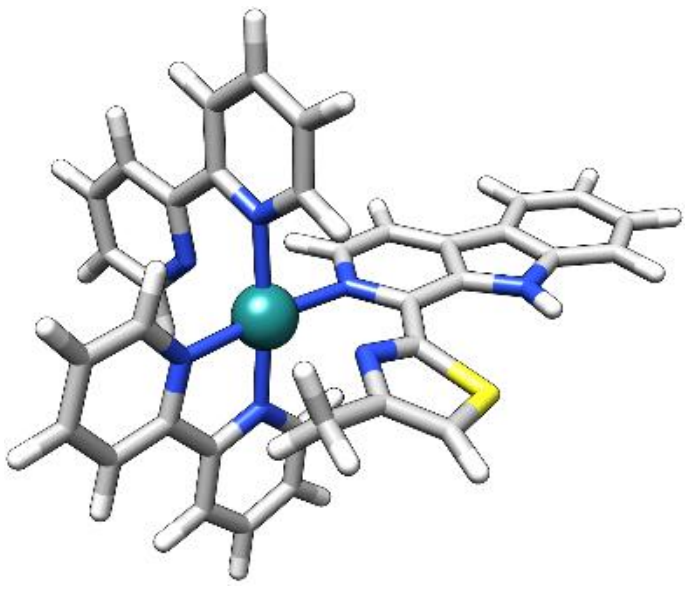<br>SOC (PBE0) = 1033.23<br>SOC (B3LYP) = 1025.85 |
| [Ru(bpy)$_2$(PhPIT)]$^{+2}$ ref. 17 | | |
| MC$_{ax\ 1}$ | MC$_{ax\ 2}$ | MC$_{ax\ 3}$ |
| 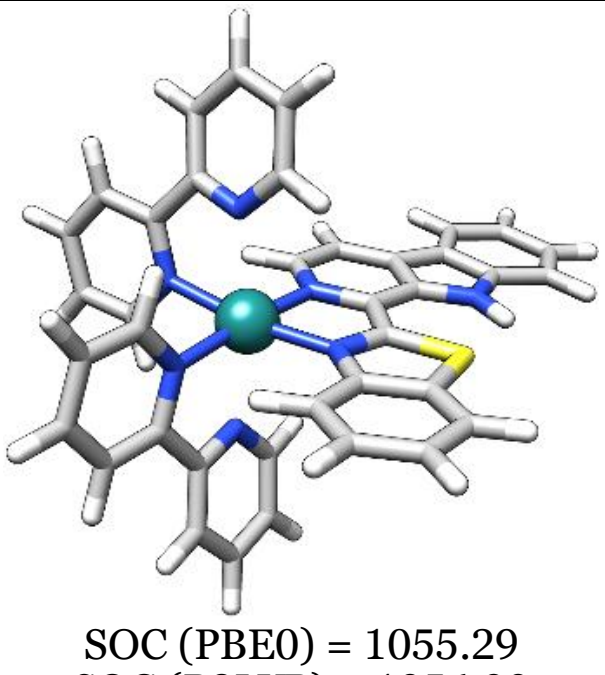<br>SOC (PBE0) = 1055.29<br>SOC (B3LYP) = 1056.90 | 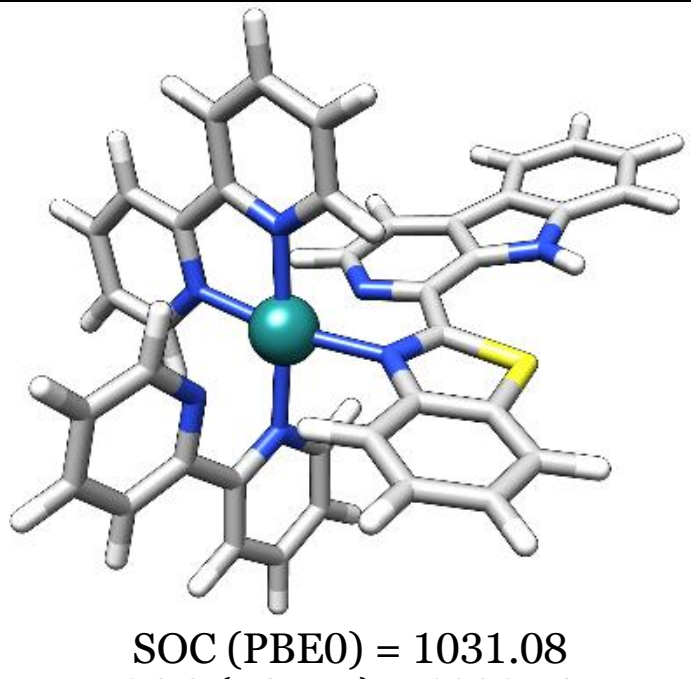<br>SOC (PBE0) = 1031.08<br>SOC (B3LYP) = 1030.73 | 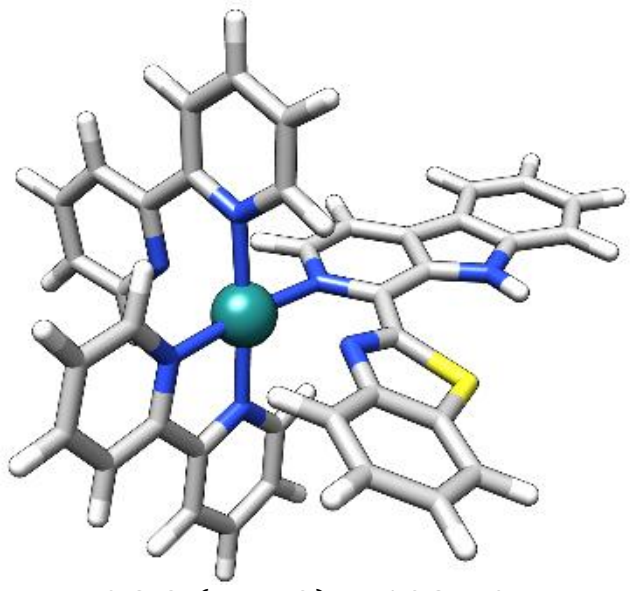<br>SOC (PBE0) = 1025.97<br>SOC (B3LYP) = 1020.46 |
| [Ru(TAP)$_2$(PIT)]$^{+2}$ ref. 17 | | |
| MC$_{ax\ 1}$ | MC$_{ax\ 2}$ | MC$_{ax\ 3}$ |
| 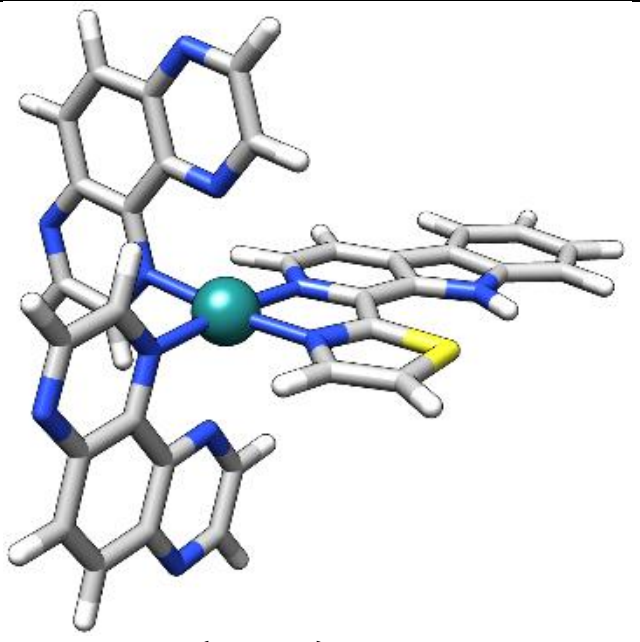<br>SOC (PBE0) = 1058.45<br>SOC (B3LYP) = 1054.32 | 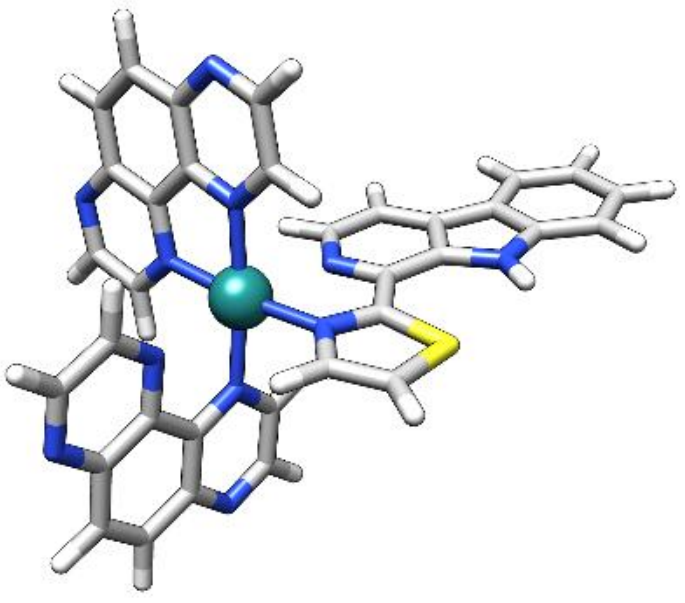<br>SOC (PBE0) = 1040.52<br>SOC (B3LYP) = 1029.33 | 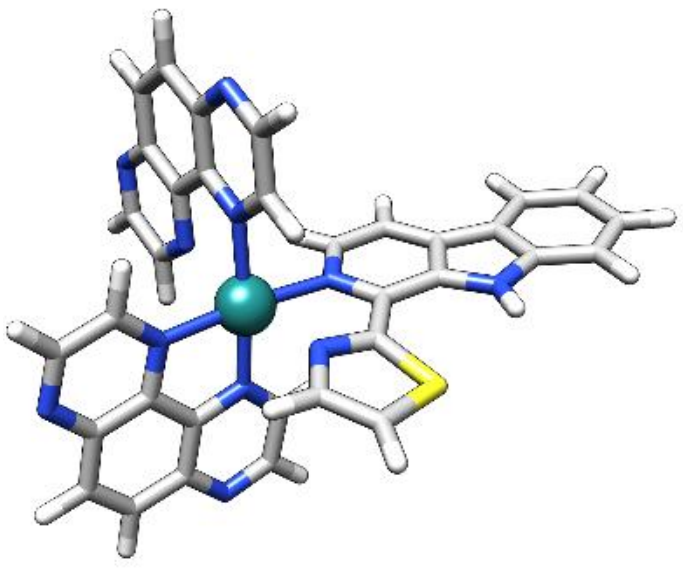<br>SOC (PBE0) = 1049.81<br>SOC (B3LYP) = 1039.42 |

| [Ru(TAP)$_2$(MePIT)]$^{+2}$ ref. 17 | | |
|---|---|---|
| MC$_{ax\,1}$ | MC$_{ax\,2}$ | MC$_{ax\,3}$ |
| 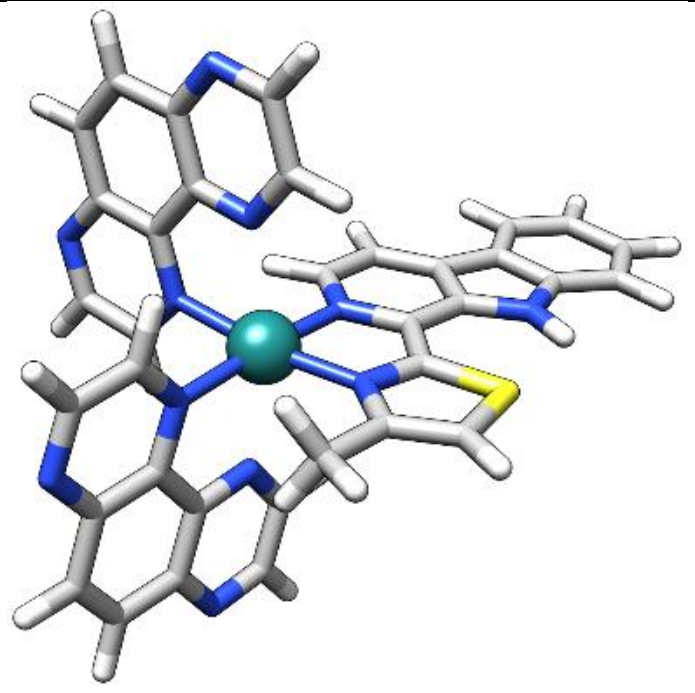<br>SOC (PBE0) = 1054.28<br>SOC (B3LYP) = 1051.60 | 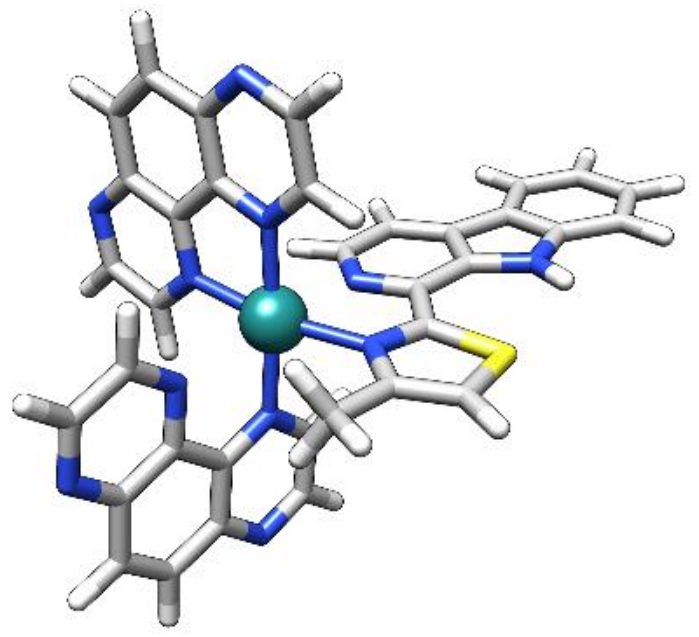<br>SOC (PBE0) = 1058.09<br>SOC (B3LYP) = 1054.79 | 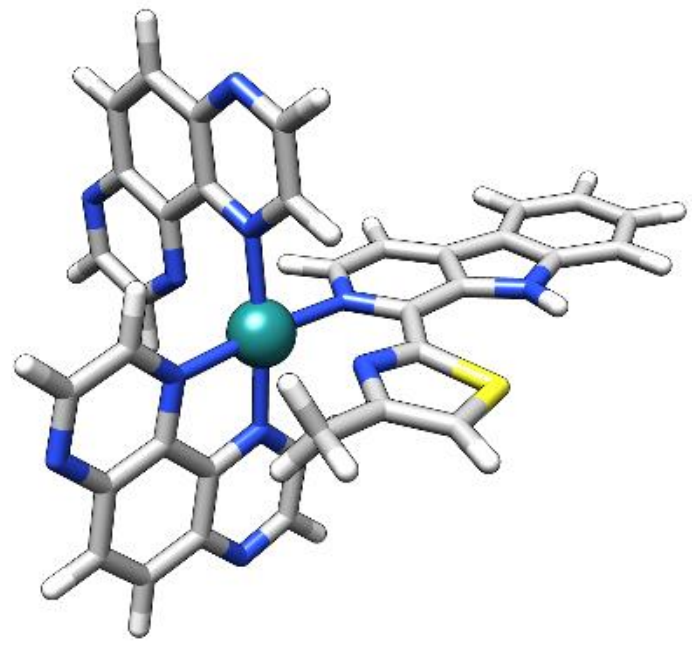<br>SOC (PBE0) = 1046.21<br>SOC (B3LYP) = 1034.79 |
| [Ru(TAP)$_2$(PhPIT)]$^{+2}$ ref. 17 | | |
| MC$_{ax\,1}$ | MC$_{ax\,2}$ | MC$_{ax\,3}$ |
| 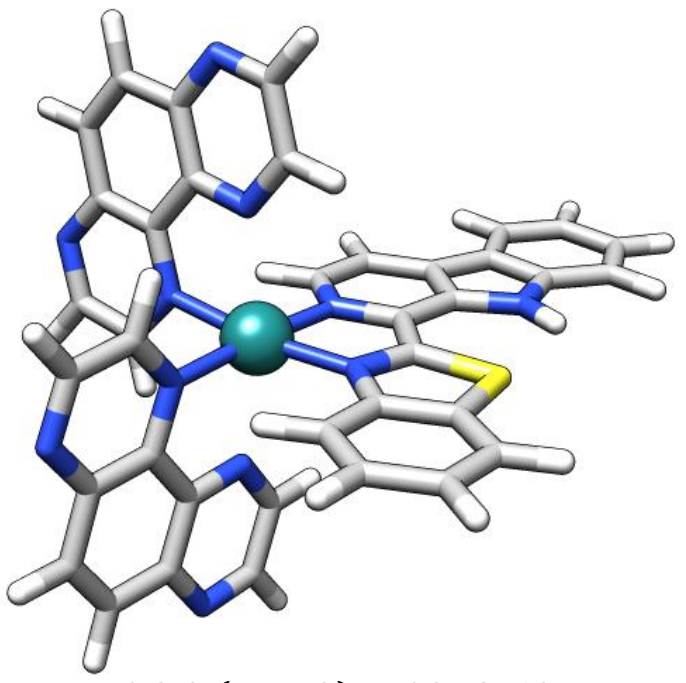<br>SOC (PBE0) = 1052.69<br>SOC (B3LYP) = 1050.88 | 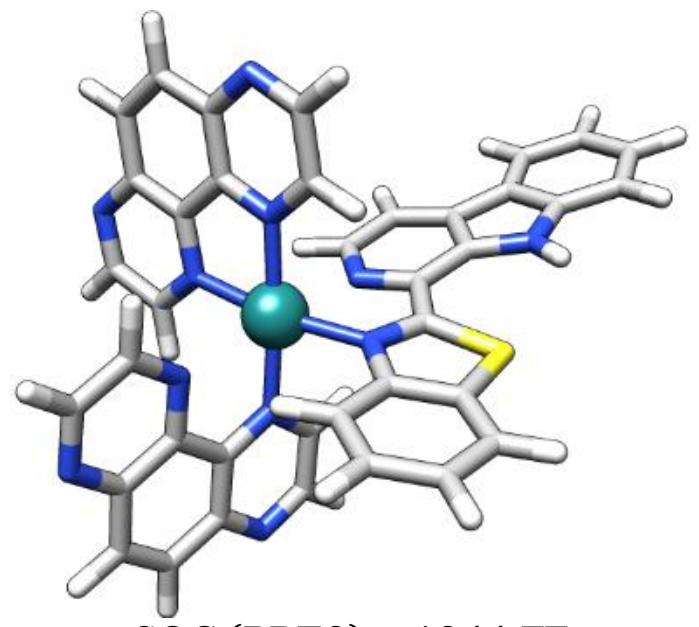<br>SOC (PBE0) = 1044.77<br>SOC (B3LYP) = 1037.49 | 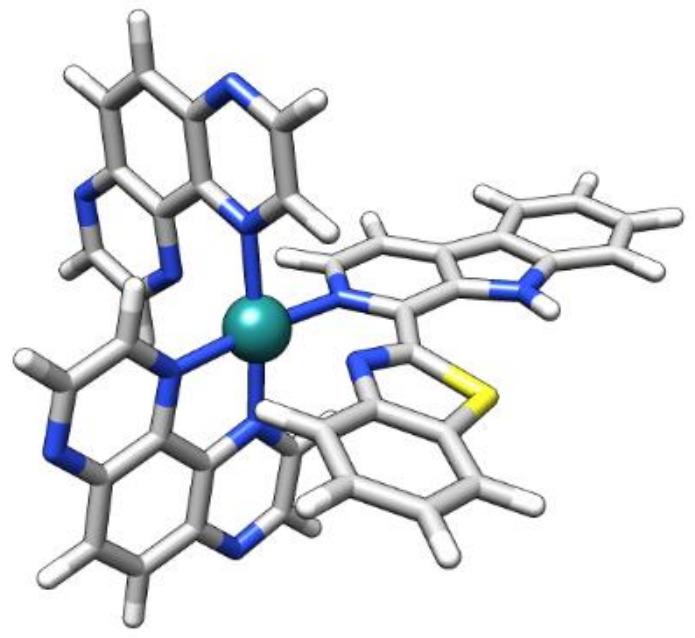<br>SOC (PBE0) = 1046.74<br>SOC (B3LYP) = 1035.11 |

**Figure S6:** Optimized geometries calculated for the $^3$MC minima of a series of Ru(II) complexes. The value of the SOC (in cm$^{-1}$) between the $^3$MC and $S_0$ states computed at the SOC-TDDFT PBE0/(ZORA-def2-TZVP + ZORA-def2-SVP level, SOC(PBE0), and at the SOC-TDDFT B3LYP/(ZORA-def2-TZVP + ZORA-def2-SVP level, SOC(B3LYP), is also reported.